\DocumentMetadata{}
\documentclass[sigconf, nonacm]{acmart}

\usepackage{listings}
\usepackage{algorithm}
\usepackage{algpseudocode}

\lstnewenvironment{SQL}
  {\lstset{
        aboveskip=5pt,
        belowskip=5pt,
        escapechar=!,
        mathescape=true,
        language=SQL,
        basicstyle=\linespread{0.94}\ttfamily,
        morekeywords={BULK, COLLECT, FOR, MATRIX, VECTOR, EACH, INTEGER, LONG, DOUBLE, WITH, PARTITION, TEST, WHETHER, PROBABILITY, VECTORIZE, ROWMATRIX, inner\_product, matrix\_vector\_multiply, MATRIX\_MULTIPLY, outer\_product, matrix\_inverse, trans\_matrix,
        label\_scalar, label\_vector, get\_scalar, diag, rowMins, rowIndexMax, map, transpose, multiply, asInstanceOf, DenseMatrix, reduce,
        toArray, zipped, toIndexedRowMatrix, if, Tuple2, override, def, new, gemm, build, filter, matmul, crossentropyderiv, reluderiv, t, relu, softmax, reducebyrow},
        deletekeywords={VALUE, PRIOR},
        showstringspaces=true}
        \vspace{0pt}%
        \noindent\minipage{0.47\textwidth}}
  {\endminipage\vspace{0pt}}
  
\newcommand\vldbdoi{XX.XX/XXX.XX}
\newcommand\vldbpages{XXX-XXX}
\newcommand\vldbvolume{14}
\newcommand\vldbissue{1}
\newcommand\vldbyear{2020}
\newcommand\vldbauthors{\authors}
\newcommand\vldbtitle{\shorttitle} 
\newcommand\vldbavailabilityurl{URL_TO_YOUR_ARTIFACTS}
\newcommand\vldbpagestyle{plain} 

\begin{document}
\title[\texttt{EinDecomp}]{\texttt{EinDecomp}: Decomposition of Declaratively-Specified Machine Learning and Numerical Computations for Parallel Execution}

\author{Daniel Bourgeois}
\affiliation{%
  \institution{Rice University}
  \streetaddress{6100 Main St}
  \city{Houston}
  \state{Texas}
  \country{United States}
  \postcode{77005}
}
\email{dcb10@rice.edu}

\author{Zhimin Ding}
\affiliation{%
  \institution{Rice University}
  \streetaddress{6100 Main St}
  \city{Houston}
  \state{Texas}
  \country{United States}
  \postcode{77005}
}
\email{zd21@rice.edu}

\author{Dimitrije Jankov}
\affiliation{%
  \institution{Rice University}
  \streetaddress{6100 Main St}
  \city{Houston}
  \state{Texas}
  \country{United States}
  \postcode{77005}
}
\email{dimitrijejankov@gmail.com}

\author{Jiehui Li}
\affiliation{%
  \institution{Rice University}
  \streetaddress{6100 Main St}
  \city{Houston}
  \state{Texas}
  \country{United States}
  \postcode{77005}
}
\email{jl302@rice.edu}

\author{Mahmoud Sleem}
\affiliation{%
  \institution{Rice University}
  \streetaddress{6100 Main St}
  \city{Houston}
  \state{Texas}
  \country{United States}
  \postcode{77005}
}
\email{msm15@rice.edu}

\author{Yuxin Tang}
\affiliation{%
  \institution{Rice University}
  \streetaddress{6100 Main St}
  \city{Houston}
  \state{Texas}
  \country{United States}
  \postcode{77005}
}
\email{yuxin.tang@rice.edu}

\author{Jiawen Yao}
\affiliation{%
  \institution{Rice University}
  \streetaddress{6100 Main St}
  \city{Houston}
  \state{Texas}
  \country{United States}
  \postcode{77005}
}
\email{jy75@rice.edu}

\author{Xinyu Yao}
\affiliation{%
  \institution{Rice University}
  \streetaddress{6100 Main St}
  \city{Houston}
  \state{Texas}
  \country{United States}
  \postcode{77005}
}
\email{xy38@rice.edu}

\author{Chris Jermaine}
\affiliation{%
  \institution{Rice University}
  \streetaddress{6100 Main St}
  \city{Houston}
  \state{Texas}
  \country{United States}
  \postcode{77005}
}
\email{cmj4@rice.edu}

\begin{abstract}
We consider the problem of automatically decomposing operations over tensors or arrays so that they can be executed in parallel on multiple devices.  We address two, closely-linked questions. First, what programming abstraction should systems for tensor-based computing offer to enable such decompositions? Second, given that abstraction, how should such systems automatically decompose a tensor-based computation?  We assert that tensor-based systems should offer a programming abstraction based on an \emph{extended Einstein summation notation}, which is a fully declarative, mathematical specification for tensor computations.  We show that any computation specified in the Einstein summation notation can be re-written into an equivalent \emph{tensor-relational} computation, and this re-write generalizes existing notations of tensor parallelism such as ``data parallel'' and ``model parallel.''  We consider the algorithmic problem of optimally computing a tensor-relational decomposition of a graph of operations specified in our extended Einstein summation notation, and we experimentally show the value of the algorithm that we develop.
\end{abstract}

\maketitle

\pagestyle{\vldbpagestyle}
\begingroup\small\noindent\raggedright\textbf{PVLDB Reference Format:}\\
\vldbauthors. \vldbtitle. PVLDB, \vldbvolume(\vldbissue): \vldbpages, \vldbyear.\\
\href{https://doi.org/\vldbdoi}{doi:\vldbdoi}
\endgroup
\begingroup
\renewcommand\thefootnote{}\footnote{\noindent
This work is licensed under the Creative Commons BY-NC-ND 4.0 International License. Visit \url{https://creativecommons.org/licenses/by-nc-nd/4.0/} to view a copy of this license. For any use beyond those covered by this license, obtain permission by emailing \href{mailto:info@vldb.org}{info@vldb.org}. Copyright is held by the owner/author(s). Publication rights licensed to the VLDB Endowment. \\
\raggedright Proceedings of the VLDB Endowment, Vol. \vldbvolume, No. \vldbissue\ %
ISSN 2150-8097. \\
\href{https://doi.org/\vldbdoi}{doi:\vldbdoi} \\
}\addtocounter{footnote}{-1}\endgroup

\ifdefempty{\vldbavailabilityurl}{}{
\vspace{.3cm}
\begingroup\small\noindent\raggedright\textbf{PVLDB Artifact Availability:}\\
The \texttt{EinDecomp} algorihtm is implemented in the \texttt{Einsummable} Machine Learning System, available at \url{https://github.com/dcbdan/einsummable}.
\endgroup
}

\section{Introduction}

Automatically partitioning numerical computations over matrices or multi-dimensional arrays (often referred to as \emph{tensors}) so that they can be run in parallel on multiple computers, cores, or compute devices such as GPUs is a key problem in modern computing. 

There are two distinct types of parallelism available to tensor-based systems.
\emph{Intra-operator} parallelism \cite{mehta1995managing,agarwal1995three,solomonik2011communication} is the ``classic'' type of parallelism, where different operators (such as different matrix multiplications) are computed in parallel on multiple sites.  While \emph{inter-operator} parallelism \cite{hasan1994optimization,narayanan2019pipedream,huang2019gpipe}  (typically realized as \emph{pipeline parallelism} in both tensor and database systems) is also important, intra-operator parallelism is typically more challenging but also (potentially) more impactful because it can reduce the latency of individual computations, such as inference in a modern large language model. In this paper, we re-write operations using \texttt{EinSum} to break the atomicity of the operator, enabling systematic optimization at a finer granularity.

\vspace{5 pt}
\noindent \textbf{Types of intra-operator parallelism.}
One of the difficulties with intra-operator parallelism is that there are a lot of ways to break up a tensor-based computation, and the various decompositions can have very different computational profiles, with the amount of communication induced being a key factor determining the ability of the decomposition to realize a parallel speedup.
Using the terminology common in machine learning (ML), the two standard methods for realizing intra-operator parallelism are \emph{data parallelism} and \emph{model parallelism}.  
\emph{Data parallelism} \cite{dean2012large,hadjis2016omnivore} is typically associated with distributed training for ML. In data parallel training, the model is replicated at each site, the data are partitioned (or \emph{sharded}) to different sites. Gradients are evaluated independently at each site, before being summed up to update the model.
In \emph{model parallelism} \cite{zhang1990efficient, farber1997parallel, raina2009large}, the model is partitioned and data are replicated.  For example, the computation of a function of $m$ outputs (such as a matrix multiplication) can typically be partitioned to $n$ sites by computing the first $m/n$ outputs at the first site, the next $m/n$ outputs at the second, and so on. 

In practice, however, data parallelism and model parallelism are not the only options \cite{zheng2022alpa,jia2019beyond}, and even the terms themselves are problematic.  In modern ML, it is not unusual to perform a series of hundreds of computations over pairs of four- or five-dimensional tensors. Deep within an ML computation, it is unclear which dimensions are ``data'' dimensions and which dimensions are ``model'' dimensions.  Further, there is no reason to assume that a partitioning may not shard along arbitrary subsets of both the data and the model dimensions.

\vspace{5 pt}
\noindent
\textbf{Programming and intra-operator parallelism.}
In modern systems for high-performance, tensor-based computations such as TensorFlow \cite{AbadiBCCDDDGIIK16} and PyTorch \cite{PaszkeGMLBCKLGA19}, the question of how to partition the operators in a complex computation is left to the programmer.  With a few notable exceptions \cite{zheng2022alpa,cai2021tensoropt,jia2019beyond,miao2022galvatron}, there has been relatively little work aimed at ``hands-free'' decomposition of tensor-based computations to enable effective, fully-automated, intra-operator parallelism.  The work that does exist does not consider the link between the programming abstraction and the decomposition.

In this paper, we address two, closely-linked questions. First, what programming abstraction should systems for tensor-based computing offer to enable intra-operator parallelism? Second, given that abstraction, how should such systems automatically decompose an tensor-based computation?  We believe that these two questions cannot be disentangled, because the selected programming abstraction needs to expose the semantics underlying the array-based computation to the system, or else the system cannot understand how to properly decompose the computation. While the programming abstraction can be hidden beneath a PyTorch-like API for programmers familiar with the current API, the current state of affairs, where operations over tensors (such as matrix multiplication) are black-box operations whose semantics are opaque to the system, cannot easily facilitate fully automated intra-operator parallelism.   Systems such as PyTorch have massive APIs, and so any ad-hoc approach that does not seek to provide a unified abstraction for specifying the semantics of computations over tensors is not likely to be practical.  As the developers of PyTorch state, ``Writing a backend for PyTorch is challenging. PyTorch has 1200+ operators, and 2000+ if you consider various overloads for each operator \cite{backend}.'' 

\vspace{5 pt}
\noindent
\textbf{Our contributions.}
This paper has three main contributions.  

First, we argue that tensor-based systems should offer a programming abstraction based on an \emph{extended Einstein summation notation} \cite{einstein1938gravitational}.  Einstein summation notation is a fully declarative, mathematical specification for tensor computations, that is common in physics and is already supported in at least some fashion by both PyToch and TensorFlow, and so it already has some buy-in. Further, it is simple and easy to understand.  If one views a tensor as a relation---tensors can be seen as relations mapping keys (lists of integers that index into the tensor) to values (some sort of scalar value)---then Einstein summation notation is closely related to other relational programming languages such as SQL, as it specifies a join followed by an aggregation over the input tensors \cite{blacher2023efficient}.  We call our expended Einstein summation notation \texttt{EinSum}.

Second, we show that any computation expressed in \texttt{EinSum} can be re-written into an equivalent \emph{tensor-relational} computation \cite{yuan2021tensor}.  A tensor-relational computation is a relational computation that operates not over relations mapping keys to scalars, but instead operates over relations mapping keys to tensors.  This equivalence is crucial because while a classical relational system operating over scalars is never going to be competitive with PyTorch or TensorFlow, tensor-relational computations are amenable to a high-performance implementation on top of a tensor-based runtime that, in theory could be implemented on top of almost any existing system for tensor computations. A tensor-relational computation pushes tensors, not scalars, through a runtime, and operates over those tensors using high-performance \emph{kernel functions} \cite{vasilache2018tensor,kjolstad2017tensor,chen2018tvm} that are carefully optimized to make full use of the CPU/GPU hardware.  

Re-writing an operation specified in \texttt{EinSum} into an equivalent, tensor-relational computation facilitates intra-operator parallelism, as it effectively decomposes the operation into a set of kernel invocations.   Tensor-relational re-writes generalize existing notions such as data parallelism and model parallelism.  
However, the space of possible decompositions is large, and further complicated when the input computation consists of many operations.  The decompositions cannot be considered independently, as decomposing an operation to support intra-operator parallelism implicitly produces a decomposition of the output tensor that may be incompatible with the required input to the next operation.

Thus, our third contribution is to consider the problem of optimally choosing a tensor-relational decomposition for a directed, acyclic graph of \texttt{EinSum} operations. We propose an algorithm, called \texttt{EinDecomp}, that does the decomposition so as to minimize the amount of communication between kernel calls in the resulting tensor-relational computation, while at the same time ensuring that there is enough work to do to keep a specified number of devices (CPU cores or GPUs) busy.

An extensive set of experiments over CPUs and GPUs shows the value of the \texttt{EinDecomp} approach.  Not only do we show that this decomposition is superior to other, heuristic-based decompositions, but we show that a system based on \texttt{EinSum} and the \texttt{EinDecomp} approach can be faster than other standard softwares, such as PyTorch \cite{PaszkeGMLBCKLGA19}, Dask \cite{rocklin2015dask}, and ScaLAPACK \cite{choi1992scalapack}.

\section{Paper Roadmap}

We begin by describing the \texttt{EinSum} language through a number of examples, including how it can be used to succinctly specify multi-headed attention \cite{vaswani2017attention} (Section \ref{sec:Einsum}). We then define the notion of a \emph{tensor relation}, which is a relation that can be used to implement a tensor as a set of keyed sub-tensors.  We describe a simple relational algebra over tensor relations called the \emph{tensor relational algebra} (TRA) that can be implemented on top of an existing runtime, like PyTorch, or as a special-purpose TRA runtime.  We then describe how any \texttt{EinSum} expression can be re-written into a computation in the TRA (Section \ref{sec:TRA}). The amount and exact nature of the parallelism available in the resulting TRA computation is controlled by a \emph{partitioning vector}.  Thus, given a complex computation (such as a large language model) specified as a graph of \texttt{EinSum} operations, the problem of decomposing it into a TRA computation is reduced to the problem of associating a partitioning vector with every computation in the graph (Section \ref{sec:Opt}). We develop a cost model for executing such a TRA computation (Section \ref{sec:Costing}), as well as a dynamic programming algorithm called \textsc{EinDecomp} that chooses the set of partitioning vectors to minimize that cost (Section \ref{sec:Eindecomp}).

\section{\texttt{EinSum} Background and Examples} \label{sec:Einsum}
We introduce the \texttt{EinSum} tensor operator by generalizing from matrix multiplication. First, we define the notion of \emph{tensor}. We use bold upper-case (for example, $\mathbf{U}$) to denote a tensor. Define the \emph{bound} vector for $\mathbf{U}$, $\mathbf{b}_{\mathbf{U}}$ to be a vector of integers of length $r$. ``r'' stands for ``rank;'' matrices are rank-2 tensors. Next, define $\mathcal{I}(\mathbf{b}_{\mathbf{U}})$ to be the set 
$\{0...\mathbf{b}_{\mathbf{U}}[0]  -1\} \times
 \{0...\mathbf{b}_{\mathbf{U}}[1]  -1\} \times ... \times
 \{0...\mathbf{b}_{\mathbf{U}}[r-1]-1\}$.
This is the set of all indices or keys that obey the bound. A tensor $\mathbf{U}$ is then a function from $\mathcal{I}({\mathbf{b}_{\mathbf{U}}})$ to the set of real numbers. 

\vspace{5 pt}
\noindent
\textbf{Simple \texttt{EinSum} examples.}
We start with the classic example: matrix multiplication.  Let $\textbf{X}$ and $\textbf{Y}$ be matrices with bounds $[100, 200]$ and $[200, 50]$, respectively.  Then matrix multiplication is written as: $\forall \hspace{2 pt} i,k \in \mathcal{I}\left([100, 50]\right)$:
\begin{align}
\textbf{Z}_{i,k} \leftarrow
\sum_{j \in \mathcal{I} \left( [200] \right)}  
  \textbf{X}_{i,j} \times \textbf{Y}_{j,k}
\end{align} \label{eqn:matmul}
For simplicity, we may drop the subscript on the aggregation operation, as it is implied; any indices that appear in the input tensors that do not appear in the output tensor must be aggregated.  We also typically drop the range specification for the output labels, as this is implied by the bound vector for the output tensor.

If instead of computing matrix multiply, we wanted to compute the squared $L^2$ distance between each row of $\mathbf{X}$ and each column of $\mathbf{Y}$, then we can replace the scalar multiplication $x \times y$ with $(x - y)^2$:
\begin{align*}
\textbf{Z}_{i,k} \leftarrow \sum (\textbf{X}_{i,j} - \textbf{Y}_{j,k})^2.
\end{align*}

To compute the $L^{\infty}$ distance instead, (i) replace $x \times y$ with $|x-y|$ and (ii) replace the summation with maximization:
\begin{align*}
\textbf{Z}_{i,k} \leftarrow \textrm{max } |\textbf{X}_{i,j} - \textbf{Y}_{j,k}|.
\end{align*}

If one views the tensors as relations mapping keys to real values, binary \texttt{EinSum} expressions perform a join of the two tensors to link values, followed by the application of a scalar function (multiplication, squared difference, etc.), followed by an aggregation. It is possible to have unary \texttt{EinSum} expressions where the join is replaced by a simple map operation that only applies a scalar function. The aggregation operator must be associative and commutative.

\vspace{5 pt}
\noindent
\textbf{General form of \texttt{EinSum}.}
In full generality, \texttt{EinSum} applies to all rank $r$ tensors, not just matrices. The general form involves more notation to specify indexing; the notation will be necessary to precisely describe decompositions of general \texttt{EinSum} expressions. 

A \emph{label} is some symbol that can be bound to a value.  We use $\ell_\textbf{U}$ to denote a list (vector) of labels used to index into tensor $\textbf{U}$. Viewed relationally, a tensor with label vector $\ell_\textbf{U}$ is equivalent to a database relation with schema $\textbf{U}(\ell_\textbf{U}[1], \ell_\textbf{U}[2], ..., \texttt{val})$. When we ``bind'' $\ell_\textbf{U}$, we are specifying specific key values we are interested in selecting for. 

Often, we will need to project or permute a bound vector. Given two lists of labels $\ell_1$ and $\ell_2$, and a bound vector $\textbf{b}$, define $\textbf{b}[\ell_1; \ell_2]$ to be a vector of length $|\ell_1|$, where the $i$th entry is $\textbf{b}[j]$ iff $\ell_1[i]$ = $\ell_2[j]$. As an example, let $\textbf{b} = [ 2, 3, 4 ]$ and let $\ell_1 = [ k, i ]$ and $\ell_2 = [ i, j, k ]$. Then $\textbf{b}[\ell_1; \ell_2] = [ 4, 2 ]$. 

Given this, binary Einsum expressions take the general form:
\begin{align}
\forall \hspace{2 pt} \ell_\textbf{Z} \in 
\mathcal{I}\left(\textbf{b}_{\textbf{Z}} \right): 
\hspace{2 pt} \textbf{Z}_{\ell_\textbf{Z}} \leftarrow \hspace{- 7 pt}
\bigoplus_{\ell_\textrm{agg} \in \mathcal{I} \left( \textbf{b}_{\textbf{X} \textbf{Y}}[\ell_{\textrm{agg}}; \ell_{\textbf{X} \textbf{Y}}] \right)} & \bigotimes \left( \textbf{X}_{\ell_\textbf{X}}, \textbf{Y}_{\ell_\textbf{Y}} \right)
\end{align} \label{eqn:Einsum}
\noindent Here, $\bigoplus$ is the aggregation operator and $\bigotimes$ is the scalar function applied to joined values (\texttt{EinSum} is an \emph{extended} Einstein summation notation as it allows for arbitrary $\bigoplus$ and $\bigotimes$ operations). In the above expression, to denote the concatenation of two label lists $\ell_\textbf{X}$ and $\ell_\textbf{Y}$, we use $\ell_\textbf{XY}$. $\textbf{b}_{\textbf{X} \textbf{Y}}$ similarly denotes the concatenation of two bound vectors.

Consider a more complicated \texttt{EinSum} expression over tensors. Assume that we have two tensors $\textbf{X}$ and $\textbf{Y}$ with bound vectors $\textbf{b}_\textbf{X} = [ 10, 100, 20 ]$ and  $\textbf{b}_\textbf{Y} = [ 100, 20, 2000 ]$.  We wish to transpose $\textbf{X}$ to obtain a tensor with bound $[ 20, 10, 100 ]$, then transpose $\textbf{Y}$ to obtain a new tensor with bound $[ 20, 100, 2000 ]$, and do a batch matrix multiply \cite{abdelfattah2020matrix} of the two resulting tensors, and then sum out the batch dimension.

In \texttt{EinSum}, this is expressed in the single expression:
\begin{align*}
\forall i,k \in \mathcal([ 10, 2000 ]), \textbf{Z}_{i,k} \leftarrow
\sum_{b,j \in \mathcal{I} \left( [ 20, 100 ] \right)}  
  \textbf{X}_{i,j,b} \times \textbf{Y}_{j,b,k}
\end{align*}
\noindent Considering the general form, we have $\ell_\textbf{X} = [ i,j,b ]$, $\ell_\textbf{Y} = [ j,b,k ]$, $\ell_\textrm{agg} = [ b,j ]$ and $\textbf{b}_\textbf{XY} = [ 10, 100, 20, 100, 20, 2000 ]$.  The bound vector for the aggregation is computed as $\textbf{b}_{\textbf{X} \textbf{Y}}[\ell_{\textrm{agg}}; \ell_{\textbf{X} \textbf{Y}}]$. How? $\ell_\textrm{agg}$ has two labels: $b$ and $j$. As $b$ occupies the third (and fifth) position in $\ell_{\textbf{X} \textbf{Y}}$, and $j$ occupies the second (and fourth) position in $\ell_{\textbf{X} \textbf{Y}}$, we select the third (or fifth) item in $\textbf{b}_\textbf{XY}$, and the second (or fourth) item results in $\textbf{b}_\textbf{XY}$. This results in the bound vector $[ 20, 100 ]$.  

When the aggregation operator is summation and the join function is multiplication, then the \texttt{EinSum} is often referred to as a \emph{contraction}. Contractions include matrix multiplication and tend to be the most computationlly challenging \texttt{EinSum} expressions. The labels that appear in inputs but not outputs are $\ell_\textrm{agg}$. If $\ell_\textrm{agg}$ is empty (meaning there are no indices being summed out), then the \texttt{EinSum} is often referred to as \emph{element-wise} and the aggregation operator may be omitted. If $\ell_\textbf{Z}$ contains labels not found in either $\ell_\textbf{X}$ or $\ell_\textbf{Y}$, then the \texttt{EinSum} is often referred to as a \emph{broadcast}, as entries are being replicated across one or more dimensions. In the remainder of the paper, we ignore broadcasts (so $\textbf{b}_\textbf{Z} = \textbf{b}_{\textbf{XY}}[\ell_\textbf{Z}; \ell_\textbf{XY}]$) and focus on contractions. We assume no repeated labels in $\ell_\textbf{X}$ or in $\ell_\textbf{Y}$, but labels are often repeated across the two sets.

\vspace{5 pt}
\noindent
\textbf{Multi-headed attention via \texttt{EinSum}.}
\texttt{EinSum} can be used to specify most modern ML computations.  For an informatative examaple, we use \texttt{EinSum} expressions to specify multi-headed attention \cite{vaswani2017attention} used by large language models. Multi-headed attention runs the attention mechanism several times in parallel, and the attention mechanism includes a softmax term.  Working backwards, we first build softmax in \texttt{EinSum}.

As applied to a vector $\mathbf{x}$, softmax gives a vector $\mathbf{y}$ where $\mathbf{y}_i = \frac{e^{\mathbf{x}_i}}{\sum_{j} e^{\mathbf{x}_j}}$. An equivalent but numerically better representation is to subtract by $c = \max_{i} \mathbf{x}_i$, giving $\mathbf{y}_i = \frac{e^{\mathbf{x}_i-c}}{\sum_{j} e^{\mathbf{x}_j-c}}$. When softmax is applied to a matrix, it is applied to each row. More generally, when the rank $r \ge 2$, softmax is applied to the last rank and ``batched'' across the first $r-1$ ranks.

For a matrix $\mathbf{X}$, softmax can be expressed in the following \texttt{EinSum}:
\begin{align*}
\textbf{C}_{i} \leftarrow \textrm{max } \textbf{X}_{i,j}                      \hspace{.4 cm}
\textbf{E}_{i,j} \leftarrow e^{\textbf{X}_{i,j} - \textbf{C}_{i}}       \nonumber \\
\textbf{S}_{i} \leftarrow \sum \textbf{E}_{i,j} \                       \hspace{.4 cm}
\textbf{Y}_{i,j} \leftarrow \frac{ \textbf{E}_{i,j} }{ \textbf{S}_{i} }
\end{align*}
As the extensions to $r > 2$ and $r = 1$ are straightforward, we assume there is an \texttt{EinSum} softmax macro that accepts any non-scalar tensor. 

The attention mechanism, applied to ``query,'' ``key'' and ``value'' matrices $\mathbf{Q}$, $\mathbf{K}$, $\mathbf{V}$, is given by $\text{softmax}(\frac{\mathbf{Q}\mathbf{K}^T}{\sqrt{d_{k}}})\mathbf{V}$, where $d_k$ is the number of columns of $\mathbf{K}$. This can be expressed as the following \texttt{EinSum}:
\begin{align*}
\textbf{T}^{(1)}_{i,k} \leftarrow \sum \textbf{Q}_{i,j} \times \textbf{K}_{k,j}        \hspace{.4 cm}
\textbf{T}^{(2)}_{i,k} \leftarrow \frac{1}{\sqrt{d_k}} \textbf{T}^{(1)}_{i,k}     \\
\textbf{T}^{(3)}       \leftarrow \text{softmax}(\textbf{T}^{(2)})                \hspace{.4 cm}
\textbf{Y}_{i,k} \leftarrow \sum \textbf{T}^{(3)}_{i,j} \times \textbf{V}_{j,k}        \hspace{.4 cm}
\end{align*}

Multi-headed attention, applied again to query, key and value matrices, is typically presented as follows:
\begin{align*}
\text{MultiHead}(\textbf{Q}, \textbf{K}, \textbf{V}) = [\textbf{H}_1, ... \textbf{H}_h] \textbf{W}_O \\
\text{where}\; \textbf{H}_i = \text{Attention}(\textbf{Q}\textbf{W}_{i}^{Q}, 
  \textbf{K}\textbf{W}_{i}^{K}
  \textbf{V}\textbf{W}_{i}^{V})
\end{align*}
Here, the input matrices are linearly projected once for each ``head'' according to weight matrices. Then, the result of each attention head is concatenated together and linearly projected to the final output space.

In the following \texttt{EinSum}, the label ``h'' stands for ``head'', ``s'' for ``sequence'' and ``a'' for ``attribute.'' First, the query, key and value matrices are linearly projected across the head dimension:
\begin{align*}
  \textbf{Q}^{H}_{s,h,d} \leftarrow \sum \textbf{Q}_{s,a} \times \textbf{W}^{Q}_{a,h,d} \\
  \textbf{K}^{H}_{s,h,d} \leftarrow \sum \textbf{K}_{s,a} \times \textbf{W}^{K}_{a,h,d} \\
  \textbf{V}^{H}_{s,h,d} \leftarrow \sum \textbf{V}_{s,a} \times \textbf{W}^{V}_{a,h,d} \\
\end{align*}
The attention computation is ``parallelized'' by batching across the head dimension:
\begin{align*}
& \textbf{T}^{(1)}_{h,s,s'} \leftarrow \sum \textbf{Q}^{H}_{s,h,d} \times \textbf{K}^{H}_{s',h,d} \hspace{.4 cm}
  \textbf{T}^{(2)}_{h,s,s'} \leftarrow \frac{1}{\sqrt{d_{k}}} \textbf{T}^{(1)}_{h,s,s'}      \\
& \textbf{T}^{(3)} \leftarrow \text{softmax}( \textbf{T}^{(2)} )                             \hspace{.4 cm}
  \textbf{O}_{s,h,d} \leftarrow \sum \textbf{T}^{(3)}_{h,s,s'} \times \textbf{V}^{H}_{s',h,d} 
\end{align*}
Lastly, the linear output projection is applied:
\begin{align*}
\textbf{Y}_{s,a} = \sum \textbf{O}_{s,h,d} \times \textbf{W}^{O}_{a,h,d} 
\end{align*}
Note that in the \texttt{EinSum} formulation, $\textbf{W}^{O}$ is not a matrix as in the standard defintion but a rank-3 tensor. This is still equivalent as the contraction producing $\textbf{Y}$ is equivalent to first concatenating the aggregation dimensions $h$ and $d$ on inputs $\textbf{O}$ and $\textbf{W}$ and then doing matrix multiply.

\section{Re-Writing \texttt{EinSum} to TRA}
\label{sec:TRA}

In this section, we describe how any computation expressed in the \texttt{EinSum} can be transformed into a computation in the tensor-relational algebra (TRA) \cite{yuan2021tensor}.  The TRA is a simple implementation abstraction that can be implemented on top of any appropriate tensor-based back-end.  Like relational algebra, TRA is trivially parallelizable, and this rewrite into TRA allows for parallelization of  \texttt{EinSum} expressions.  Translating \texttt{EinSum} into TRA so it can be executed on a tensor backend is analogous to translating SQL into relational algebra so it can be implemented on a database backend.

\subsection{Tensor Relations}

The TRA operates over \emph{tensor relations}.  A tensor relation may be viewed as a set of pairs of the form
$$(\texttt{key}, \texttt{tensor})$$  Mathematically, it is a function mapping keys to tensors.  
Like a tensor, a tensor relation $\mathcal{R}$ has a bound vector $\textbf{b}_\mathcal{R}$, but it also has a \emph{partitioning vector} $\textbf{d}_\mathcal{R}$.  The tensor relation is then a function $$\mathcal{R} : \mathcal{I} \left(\textbf{d}_\mathcal{R}\right) \rightarrow \left(\mathcal{I}\left(\frac{\textbf{b}_\mathcal{R}} {\textbf{d}_\mathcal{R}}\right) \rightarrow \mathbb{R}\right)$$
\noindent Here, the division $\frac{\textbf{b}_\textbf{R}} {\textbf{d}_\textbf{R}}$ operates element-wise over the vectors.  Thus, we can evaluate $\mathcal{R}$ at any value $\textbf{i} \in \mathcal{I} \left(\textbf{d}_\mathcal{R}\right)$ and obtain a tensor.  We use $\mathcal{R}^\textbf{i}$ to denote this evaluation.  For a tensor $\textbf{R}$, we use $\textbf{R}_\textbf{j}$ to denote the real number obtained when evaluating $\textbf{R}$ at $\textbf{j}$.  When we index a tensor relation to select a particular sub-tensor, and then we index into the sub-tensor to select a scalar, we write $\mathcal{R}^\textbf{i}_\textbf{j}$.

We say that a tensor $\textbf{R}$ and a tensor relation $\mathcal{R}$ having the same bound are \emph{equivalent}, denoted $\textbf{R} \equiv \mathcal{R}$, if, for all vectors $\textbf{j} \in \mathcal{I} \left(\textbf{d}_\textbf{R}\right)$:
$$\textbf{R}_\textbf{j} = \mathcal{R}^{\frac{\textbf{j} }{\textbf{d}_\mathcal{R}}}_{\textbf{j} \textrm{ mod }\textbf{d}_\mathcal{R}}$$
\noindent 
Again, in the above definition we assume the ``mod'' operation operates element-wise over the input vectors. $\textbf{R} \equiv \mathcal{R}$ implies that the tensor and tensor relation are alternative implementations for the same function.  Intuitively, we say that $\textbf{R} \equiv \mathcal{R}$ when $\mathcal{R}$ stores $\textbf{R}$, broken into a set of sub-tensors.

This may seem quite abstract, so it is best to illustrate this with an example.  Consider the matrix \textbf{U}: 
$$\textbf{U} = \begin{bmatrix}
1 & 2 & 5 & 6 \\
3 & 4 & 7 & 8 \\
9 & 10 & 13 & 14 \\
11 & 12 & 15 & 16
\end{bmatrix}.$$
\noindent As this is a $4 \times 4$ matrix, the bound vector $\textbf{b}_\textbf{U} = [ 4, 4 ]$.  Now, let us  imagine that we instead wanted to represent $\textbf{U}$ as a tensor relation $\mathcal{U}$ where $\textbf{b}_\mathcal{U} = [ 4, 4 ]$ and $\textbf{d}_\mathcal{U} = [ 4, 2 ]$. This $\textbf{d}_\mathcal{U}$ implies that we slice the first dimension 4 ways, and the second two ways, so if $\textbf{U} \equiv \mathcal{U}$, then
\begin{align} \mathcal{U} = \Bigg\{& \left( \langle 0, 0 \rangle, \begin{bmatrix} 1 \\ 3 \end{bmatrix}  \right),
        \left( \langle 0, 1 \rangle, \begin{bmatrix} 2 \\ 4\end{bmatrix}  \right), \left( \langle 0, 2 \rangle, \begin{bmatrix} 5 \\ 7 \end{bmatrix}  \right), \left( \langle 0, 3 \rangle, \begin{bmatrix} 6 \\ 8 \end{bmatrix}  \right), \nonumber \\
        &\left( \langle 1, 0 \rangle, \begin{bmatrix} 9 \\ 11 \end{bmatrix}  \right),
        \left( \langle 1, 1 \rangle, \begin{bmatrix} 10 \\ 12\end{bmatrix}  \right), \left( \langle 1, 2 \rangle, \begin{bmatrix} 13 \\ 15 \end{bmatrix}  \right), \left( \langle 1, 3 \rangle, \begin{bmatrix} 14 \\ 16 \end{bmatrix}  \right) \nonumber
\Bigg\}.\end{align}
This can be viewed as a function from $\mathcal{I}([ 4, 2])$ to sub-tensors with bound vector $\frac{[ 4, 4 ]}{[ 4, 2 ]} = [ 1, 2 ]$.  

If instead we let $\textbf{d}_\mathcal{U} = [ 2, 2 ]$, then we slice both dimensions two ways, and so if $\textbf{U} \equiv \mathcal{U}$, then \begin{align} \mathcal{U} = \Bigg\{& \left( \langle 0, 0 \rangle, \begin{bmatrix} 1 & 2 \\ 3 & 4 \end{bmatrix}  \right),
        \left( \langle 0, 1 \rangle, \begin{bmatrix} 5 & 6 \\ 7 & 8 \end{bmatrix}  \right), \nonumber \\
        &\left( \langle 1, 0 \rangle, \begin{bmatrix} 9 & 10 \\ 11 & 12 \end{bmatrix}  \right),
        \left( \langle 1, 1 \rangle, \begin{bmatrix} 13 & 14 \\ 15 & 16 \end{bmatrix}  \right) \nonumber
\Bigg\}.\end{align}
Effectively, to create a tensor relation $\mathcal{U}$ that is equivalent to $\textbf{U}$, we simply slice up $\textbf{U}$ according to the partitioning vector $\textbf{d}_\mathcal{U}$, and represent $\textbf{U}$ as a set of keyed sub-tensors.

\subsection{The Tensor-Relational Algebra}

The TRA is an algebra over tensor relations, that can serve as the implementation abstraction or interface that is exported by a high-performance runtime. It is closely related to the classic relational algebra. As we will show, the TRA can easily be used to implement \texttt{EinSum}.  The three TRA operations we are concerned with are \emph{join}, \emph{aggregation}, and \emph{repartition}.  

\vspace{5 pt}
\noindent
\textbf{Join.}
Given two tensor relations $\mathcal{X}$ and $\mathcal{Y}$, join applies a kernel function $K$ to each pair of sub-tensors from the two inputs that ``match''. Assume that we have a  tensor-valued function $K$ accepting two sub-tensors having bounds $\frac{\textbf{b}_\mathcal{X}}{\textbf{d}_\mathcal{X}}$ and $\frac{\textbf{b}_\mathcal{Y}}{\textbf{d}_\mathcal{Y}}$. We have two label vectors $\ell_\mathcal{X}$ and $\ell_\mathcal{Y}$.  Then $\Join_{K, \ell_\mathcal{X}, \ell_\mathcal{Y}}(\mathcal{X}, \mathcal{Y})$ first joins  $\mathcal{X}$ and $\mathcal{Y}$, matching $x \in \mathcal{X}$ and $y \in \mathcal{Y}$ iff $x.\texttt{key}[i]$ = $y.\texttt{key}[i]$ whenever $\ell_\mathcal{X}[i] = \ell_\mathcal{Y}[j]$. When $x$ and $y$ match, a new tuple $t$ is included in the output.  $t.\texttt{key}$ is $x.\texttt{key}$ concatenated with $y.\texttt{key}$ (with redundant dimensions that are trivially equal, per the join predicate, removed, as in a natural join).  $t.\texttt{tensor}$ is $K (x.\texttt{tensor}, y.\texttt{tensor})$.

\vspace{5 pt}
\noindent
\textbf{Aggregation.}  Given a tensor relation $\mathcal{X}$, aggregation performs a grouping of the tensor relation and then applies a commutative and associative kernel function $\oplus$ to reduce the values in each group to a single value. Assume that we have a  tensor-valued function $\oplus$ accepting two tensors both having bounds $\frac{b_\mathcal{X}}{d_\mathcal{X}}$.  Assume we have two label vectors $\ell_\mathcal{X}$ and $\ell_\textrm{agg}$.  
$\sum_{\oplus,\ell_\mathcal{X},\ell_\textrm{agg}}(\mathcal{X})$ first partitions $\mathcal{X}$ so two tuples $x_1$ and $x_2$ are in the same partition iff
$x_1.\texttt{key}[i]$ = $x_2.\texttt{key}[i]$ whenever $\ell_\mathcal{X}[i]$ is not in $\ell_\textrm{agg}$. 
For each partition, produce an output tuple $t$ by picking a tuple $x$ from the partition, and setting $t.\texttt{key}[i] = x.\texttt{key}[j]$ whenever $\ell_\textrm{agg}[i] = \ell_\mathcal{X}[j]$.  
$t.\texttt{tensor}$ is produced by reducing all of the \texttt{tensor} values in the partition using $\oplus$.  Note that if  $\ell_\mathcal{X}$ and $\ell_\textrm{agg}$ are identical, the aggregation is the identity operation because there are no labels in $\ell_\mathcal{X}$ not in $\ell_\textrm{agg}$.

\vspace{5 pt}
\noindent
\textbf{Repartition.} Given a tensor relation $\mathcal{X}$, let $\textbf{X} \equiv \mathcal{X}$. Then
$\Pi_{\textbf{d}} (\mathcal{X})$ produces the tensor relation $\mathcal{X}'$ such that $\textbf{d}_{\mathcal{X}'} = \textbf{d}$ and $\textbf{X} \equiv \mathcal{X}'$.

\subsection{\texttt{EinSum} As a Tensor-Relational Programming Language}

We now show that any binary \texttt{EinSum} expression can be converted to an equivalent tensor-relational computation. The benefit is that such a rewrite allows \texttt{EinSum} expressions to be run in parallel, on a set of devices.  This is done via the introduction of a partition vector that ``explodes'' the \texttt{EinSum} computation into a TRA expression over tensor relations, subsuming traditional notions such as ``data parallel'' and ``model parallel.'' 

\begin{figure}[t]
\includegraphics[width=8.5cm]{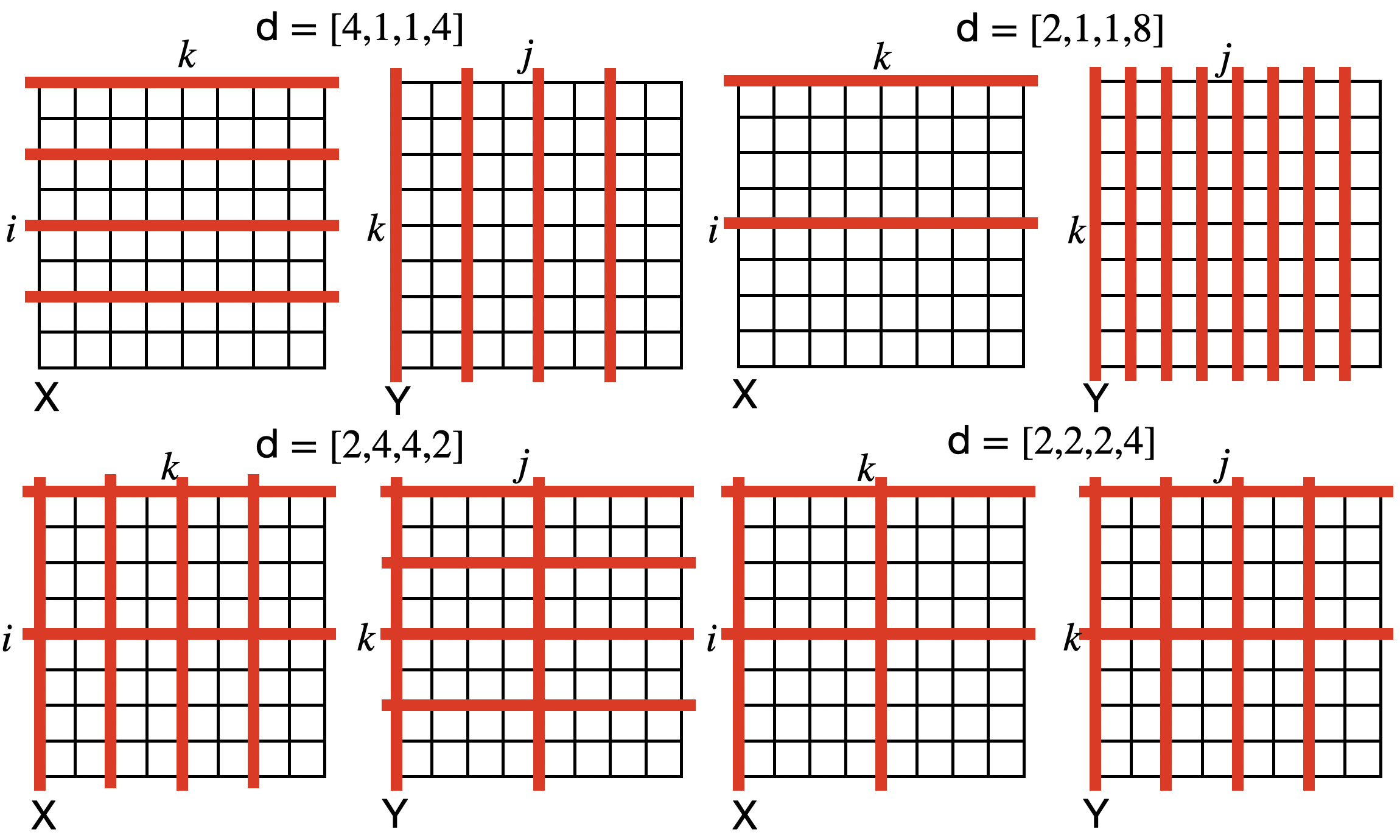}
\vspace{-15 pt}
 \caption{Four tensor-relational partitionings for  $\textbf{Z}_{i,k} \leftarrow \sum \textbf{X}_{i,j} \times \textbf{Y}_{j,k}$. In each there are 16 kernel calls. }
 \vspace{-5 pt}
 \label{fig:part}
\end{figure}

The rewrite relies on the idea of ``cloning'' each list of labels in a label list such as $\ell_\textbf{U}$ to produce a new label list $\bar{\ell}_\textbf{U}$.  The original list of labels iterates through the tuples in the tensor relation, and then the cloned list iterates through the entries in a tensor stored in the tensor relation.  For example, we may take the matrix multiplication from Equation \ref{eqn:matmul} and rewrite it as
$\forall \hspace{2 pt} i,k \in \mathcal{I}\left([2, 2]\right), \forall \hspace{2 pt} \bar{i},\bar{k} \in \mathcal{I}\left([50, 25]\right)$:
\begin{align}
\textbf{Z}_{\substack{i\times50 + \bar{i}, \\ k\times25 + \bar{k}}} \leftarrow
\sum_{\substack{j \in \mathcal{I} \left( [2] \right), \\ \bar{j} \in \mathcal{I} \left( [100] \right)}} 
  \textbf{X}_{\substack{i\times50 + \bar{i},\\ j\times100 + \bar{j}}} \times \textbf{Y}_{\substack{j\times100 + \bar{j}, \\ k\times25 + \bar{k}}} \nonumber
\end{align}
\noindent We can generalize this idea to any \texttt{EinSum} expression of the form given as Equation \ref{eqn:Einsum}. Assume we are given a partition vector $\textbf{d}$.  We can now rewrite Equation \ref{eqn:Einsum} as:
\begin{align}
&\forall \ell_\textbf{Z} \in 
\mathcal{I}\left(\textbf{d}\left[\ell_{\textbf{Z}}; \ell_{\textbf{X} \textbf{Y}}\right]\right), \bar{\ell}_\textbf{Z} \in 
\mathcal{I} \left( \frac{\textbf{b}_{\textbf{X} \textbf{Y}}} {\textbf{d}} \left[ \ell_{\textbf{Z}}; \ell_{\textbf{X} \textbf{Y}} \right] \right): 
 \textbf{Z}_{\ell_\textbf{Z} \times \frac{\textbf{b}_{\textbf{X} \textbf{Y}}} {\textbf{d}}\left[\ell_{\textbf{Z}}; \ell_{\textbf{X} \textbf{Y}} \right] + \bar{\ell}_\textbf{Z} } \nonumber 
\\
&\leftarrow  \bigoplus_{\substack{\ell_\textrm{agg} \in \mathcal{I} \left( \textbf{d} [ \ell_{\textrm{agg}} ; \ell_{\textbf{X} \textbf{Y}}] \right) \\ \bar{\ell}_\textrm{agg} \in \mathcal{I} \left( \frac{\textbf{b}_{\textbf{X} \textbf{Y}}} {\textbf{d}} [ \ell_{\textrm{agg}} ; \ell_{\textbf{X} \textbf{Y}}] \right)}}
\hspace{-10 pt} \bigotimes \left( \textbf{X}_{\ell_\textbf{X} \times \textbf{d} [\ell_\textbf{X}, \ell_\textbf{XY}] + \bar{\ell}_\textbf{X}},
\textbf{Y}_{\ell_\textbf{Y} \times \textbf{d} [\ell_\textbf{Y}; \ell_\textbf{XY}] + \bar{\ell}_\textbf{Y}} \right)  \label{eqn:TR}
\end{align}
Now, assume that we have two tensor relations $\mathcal{X} \equiv \textbf{X}$, and $\mathcal{Y} \equiv \textbf{X}$, where $\textbf{d}_\mathcal{X} = \textbf{d}[\ell_\textbf{X}; \ell_\textbf{XY}]$ and $\textbf{d}_\mathcal{Y} = \textbf{d}[\ell_\textbf{Y}; \ell_\textbf{XY}]$.  We can rewrite Equation \ref{eqn:TR} to compute a tensor relation $\mathcal{Z} \equiv \textbf{Z}$ with $\textbf{d}_\mathcal{Z} = \textbf{d}[\ell_\textbf{Z}; \ell_\textbf{XY}]$ as follows:
\begin{align}
\forall \ell_\textbf{Z} \in 
\mathcal{I}\left(\textbf{d}\left[\ell_{\textbf{Z}}; \ell_{\textbf{X} \textbf{Y}}\right]\right), \bar{\ell}_\textbf{Z} \in 
\mathcal{I} \left( \frac{\textbf{b}_{\textbf{X} \textbf{Y}}} {\textbf{d}} \left[ \ell_{\textbf{Z}}; \ell_{\textbf{X} \textbf{Y}} \right] \right): 
\mathcal{Z}^{\ell_\textbf{Z}}_{\bar{\ell}_\textbf{Z}} \nonumber 
\\
\leftarrow  \bigoplus_{\substack{\ell_\textrm{agg} \in \mathcal{I} \left( \textbf{d} [ \ell_{\textrm{agg}} ; \ell_{\textbf{X} \textbf{Y}}] \right) \\ \bar{\ell}_\textrm{agg} \in \mathcal{I} \left( \frac{\textbf{b}_{\textbf{X} \textbf{Y}}} {\textbf{d}} [ \ell_{\textrm{agg}} ; \ell_{\textbf{X} \textbf{Y}}] \right)}}
\bigotimes \left( \mathcal{X}^{\ell_\textbf{X}}_{\bar{\ell}_\textbf{X}},
\mathcal{Y}^{\ell_\textbf{Y}}_{\bar{\ell}_\textbf{Y}} \right) \label{eqn:TR2}
\end{align} 
Note that we effectively have a nested \texttt{EinSum} expression: the outer one is operating over tensor relations, and the inner one is operating over tensors in those two tensor relations.

\begin{figure*}[th!]
\includegraphics[width=18cm]{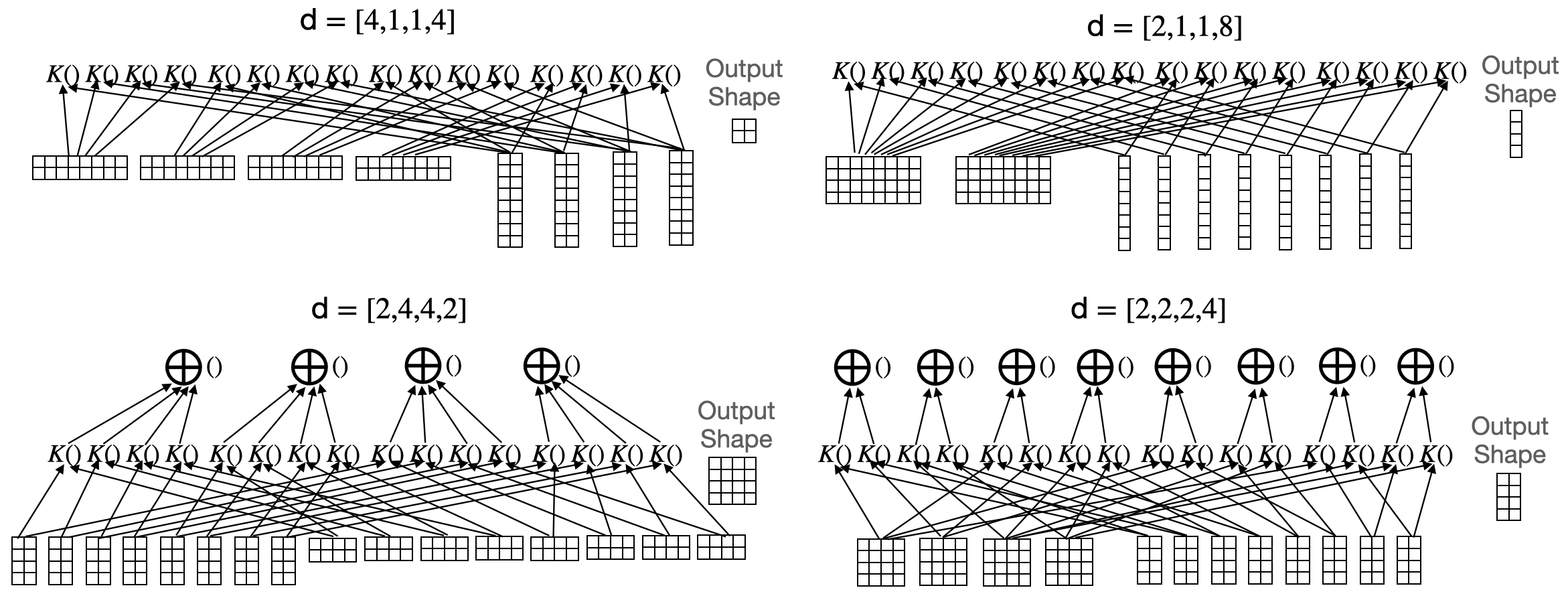}
\vspace{-18 pt}
 \caption{Dataflow graphs associated with the partitionings of Figure \ref{fig:part} For partitionings $\textbf{d} = [4, 1, 1, 4]$ and  $\textbf{d} = [2, 1, 1, 8]$, there is only a join layer, as the joined dimensions are not partitioned.  For $\textbf{d} = [2, 4, 4, 2]$ and  $\textbf{d} = [2, 2, 2, 4]$ there is also an aggregation. }
 \vspace{-5 pt}
 \label{fig:dataflow}
\end{figure*}

Now, imagine that we have a \emph{kernel function} $K$ that accepts two sub-tensors $\mathcal{X}^{\ell_\textbf{X}}$ and $\mathcal{Y}^{\ell_\textbf{Y}}$ and creates a tensor $\textbf{Z}'$ with bound vector $\frac{\textbf{b}_{\textbf{X} \textbf{Y}}} {\textbf{d}}$  such that $\forall \bar{\ell}_\textbf{Z} \in 
\mathcal{I} \left( \frac{\textbf{b}_{\textbf{X} \textbf{Y}}} {\textbf{d}} \left[ \ell_{\textbf{Z}}; \ell_{\textbf{X} \textbf{Y}} \right] \right)$,
$\textbf{Z}'_{\bar{\ell}_\textbf{Z}}$

\noindent is set to 
$\bigoplus_{ \bar{\ell}_\textrm{agg} \in \mathcal{I} \left( \frac{\textbf{b}_{\textbf{X} \textbf{Y}}} {\textbf{d}} [ \ell_{\textrm{agg}} ; \ell_{\textbf{X} \textbf{Y}}] \right)}
\bigotimes \left( \mathcal{X}^{\ell_\textbf{X}}_{\bar{\ell}_\textbf{X}},
\mathcal{Y}^{\ell_\textbf{Y}}_{\bar{\ell}_\textbf{Y}} \right)$.   If $\bigoplus$ over two tensors performs the $\bigoplus$ operation element-wise over entries in the tensors, then, using this kernel, Equation \ref{eqn:TR2} becomes:
\begin{align}
\forall \ell_\textbf{Z} \in 
\mathcal{I}\left(\textbf{d}\left[\ell_{\textbf{Z}}; \ell_{\textbf{X} \textbf{Y}}\right]\right): 
\mathcal{Z}^{\ell_\textbf{Z}} \leftarrow \hspace{-10 pt} \bigoplus_{\ell_\textrm{agg} \in \mathcal{I} \left( \textbf{d} [ \ell_{\textrm{agg}} ; \ell_{\textbf{X} \textbf{Y}}] \right)}
K \left( \mathcal{X}^{\ell_\textbf{X}},
\mathcal{Y}^{\ell_\textbf{Y}} \right)
\end{align}
\noindent The above equating is then implemented in the TRA by the two steps; a join to link tuples from $\mathcal{X}$ and $\mathcal{X}$ using the label lists $\ell_\mathcal{X}$ and $\ell_\mathcal{Y}$ followed by an aggregation using $\oplus$:
\begin{align}
\texttt{temp} \leftarrow \Join_{K, \ell_\mathcal{X}, \ell_\mathcal{Y}}(\mathcal{X}, \mathcal{Y}) \nonumber \\
\texttt{res} \leftarrow \sum_{\oplus,\ell_\mathcal{X} \odot  \ell_\mathcal{Y},\ell_{\textrm{agg}}}(\texttt{temp}) \nonumber
\end{align}
In the above, $\odot$ concatenates the two label lists, removing any duplicate labels in the process (as in the schema output from a classic natural join).

\subsection{Parallelism via the Partitioning Vector}

Thus, we have shown that any binary \texttt{EinSum} expression can be re-written as an equivalent computation in the TRA, that can then be implemented by a TRA runtime. Crucially, the way that the \texttt{EinSum} expression is decomposed into tensor relations is controlled by the partitioning vector $\textbf{d}$; choosing a different $\textbf{d}$ produces different partitionings of the input tensors, ``exploding'' the computations in different ways.

For example, the matrix multiplication of two $8 \times 8$ matrices, specified via $\textbf{Z}_{i,k} \leftarrow \sum \textbf{X}_{i,j} \times \textbf{Y}_{j,k}$.  The decomposition of the input matrices, as well as the resulting TRA computation, are fully specified with a $\textbf{d}$ vector having four entries (because both inputs are rank two).  Figure \ref{fig:part} depicts how four different $\textbf{d}$ vectors specify four different decompositions of the two input matrices.  Figure \ref{fig:dataflow} depicts the four associated dataflow (or lineage) graphs.  These graphs show how the TRA computations associated with the decompositions shown in Figure \ref{fig:part} map tuples into kernel calls.

\section{Optimizing the Decomposition}
\label{sec:Opt}

A complex computation specified in \texttt{EinSum} can be represented as a directed, acyclic graph called an \textsc{EinGraph} whose nodes are \texttt{EinSum} expressions and whose edges represent data flow.  For each vertex, we have a triple: 
$$(\texttt{bound}, \texttt{EinSum}, \texttt{inputs})$$
\texttt{EinSum} is the code run at the vertex, \texttt{bound} is the bound vector $\textbf{b}$ for the output of the \texttt{EinSum}, and \texttt{inputs} lists vertices providing inputs to this vertex.  Note that \texttt{inputs} has an explicit ordering, as \texttt{EinSum} need not be commutative.  \texttt{inputs} is empty if and only if \texttt{EinSum} empty; then the tensor is input to the computation.  Otherwise, the \texttt{bound} can be deduced from the \texttt{EinSum} labels and the input tensor shapes.

As described in the previous section, an \textsc{EinGraph} can be executed by a TRA engine. Once a partition vector describing each \texttt{EinSum} operation is to be parallelized has been associated with each vertex in the graph, each vertex is executed as a join followed by an aggregation, with (optionally) a repartition required in between operations if the output partitioning of an operation does not match the required input partitioning of the next operation.

\begin{figure}[t]
\includegraphics[width=8.5cm]{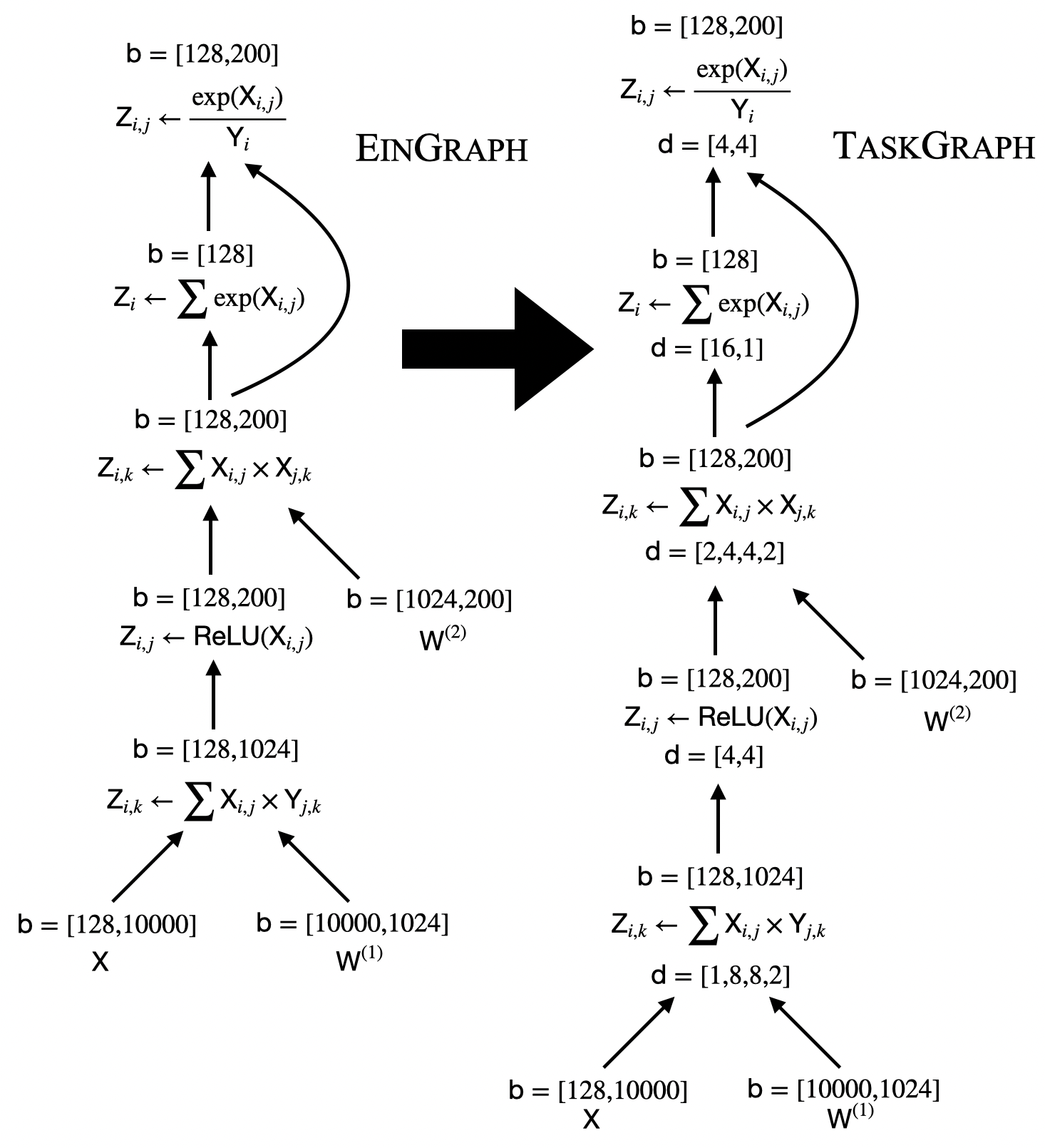}
\vspace{-23 pt}
 \caption{Modifying an \textsc{EinGraph} supplied by a programmer, to produce a \textsc{TaskGraph}, by adding bound vectors.  }
 \vspace{-12 pt}
 \label{fig:optlabel}
\end{figure}

A key question is: How to associate a partitioning vector with each operation in the \textsc{EinGraph}, as this can have a radical impact on system performance?  Thus, we now consider: Given an \textsc{EinGraph}, how do we annotate the \textsc{EinGraph} with partitioning vectors (see Figure \ref{fig:optlabel}) to describe how the inputs each vertex in the \textsc{EinGraph} are to be partitioned, to produce the best partitioning?

There will be two key considerations when labeling the \textsc{EinGraph} with partition vectors.  First, we want to ensure that there is enough parallel work to do.  And second, we want to ensure that the decomposition we choose is low cost.  The question of how much parallel work is associated with a decomposition, and how to cost a decomposition, are considered in the next two sections of the paper.  

\section{Ensuring Enough Parallel Work}
\label{sec:Parallel}

Assume the underlying system has $p$ ``processors'' (in practice these may be CPU cores, GPUs, or FPGAs). We want to ensure that the tensor relational implementation of each \texttt{EinSum} expression is decomposed to at least $p$ independent calls to the kernel $K$.  At the same time, we do not want \emph{too many} independent kernel calls, as more kernel calls tend to induce more movement across processors.  Thus, we attempt to decompose each \texttt{EinSum} expression into exactly $p$ kernel calls.

How to ensure this?  A partitioning vector $\textbf{d}$ controls the decomposition of an \texttt{EinSum} computation. $\textbf{d}$ partitions a tensor $\textbf{d}[i]$ ways along the $i$th dimension.  
For a binary \texttt{EinSum} expression over tensors $\textbf{Z}_{\ell_\textbf{Z}} \leftarrow \bigoplus \bigotimes \left( \textbf{X}_{\ell_\textbf{X}}, \textbf{Y}_{\ell_\textbf{Y}} \right)$, $\textbf{d}$ partitions \textbf{X} according to $\textbf{d}[\ell_\textbf{X} ; \ell_\textbf{XY}]$ and \textbf{Y} according to $\textbf{d}[\ell_\textbf{Y} ; \ell_\textbf{XY}]$.  Note that the elements in $\textbf{d}$ corresponding to the same label must be the same.  For example, Figure 1 shows four possible partitioning vectors (and the corresponding partitionings) for a matrix multiply $\textbf{Z}_{i,k} \leftarrow \sum \textbf{X}_{i,j} \times \textbf{Y}_{j,k}$. Here, valid partitionings will have $\mathbf{d}[1] = \mathbf{d}[2]$, corresponding to the shared $j$ label.

When an \texttt{EinSum} expression is partitioned according to $\textbf{d}$, the number of pairs of tuples matched during the join is $N (\ell_\textbf{X}, \ell_\textbf{Y}, \textbf{d}) =  \prod \textbf{d}[\ell_\textbf{X} \odot \ell_\textbf{Y}; \ell_\textbf{XY}]$.  Recall that $\odot$ concatenates the two label lists, removing any duplicate labels in the process.
What is the intuition behind this formula? Repeated labels correspond to an equality predicate associated with the join, and an equality predicate cuts down the number of tuples resulting from the join by a factor of $d$ when the number of partitions along the corresponding dimension is $d$.  So, in our matrix multiply example, $\textbf{d} = [16, 2, 2, 4]$ would result in $16 \times 2 \times 4 = 128$ tuples being output from the join; as the second $2$ is associated with a join predicate and does not contribute join results.

The series of dataflow graphs shown in Figure \ref{fig:dataflow} depict how tuples are linked and kernel functions invoked, for four tensor-relational implementations of the matrix multiplication $\textbf{Z}_{i,k} \leftarrow \sum \textbf{X}_{i,j} \times \textbf{Y}_{j,k}$.  Note that for each of these partition vectors, $N ([i, j], [j, k], \textbf{d}) = 16$.

\section{Costing A Decomposition}
\label{sec:Costing}

Our approach to costing a decomposition is to compute an upper bound on the number of floating point numbers that must be transferred to implement the resulting tensor relational computation, and to use that as the cost.  That is, we assume the worst, that every input to a node in the dataflow graph must be transferred to the processor where it is to be used.  Counting transfers makes sense as all decompositions will have the same total number of floating point operations.

To execute a vertex in an \textsc{EinGraph}, there are three steps that incur data transfer, corresponding to the join, aggregation, and possible repartition of the output that are necessary to implement the vertex:

\begin{enumerate}
    \item Transferring sub-tensors to where two tuples are to be joined.
    \item Transferring sub-tensors resulting the join to the location where they are to be aggregated.
    \item Possibly re-partitioning the result of the aggregation to be used in subsequent \textsc{EinGraph} nodes.
\end{enumerate}

\noindent We now consider how to cost each of these.

\vspace{5 pt}
\noindent
\textbf{Transferring into the join.}  Let $n_{\textbf{X}}$ be the count of floating point numbers in each sub-tensor from $\textbf{X}$. This is  
$\prod \frac{\textbf{b}_{\textbf{X} \textbf{Y}}} {\textbf{d}} [ \ell_{\textbf{X}} ; \ell_{\textbf{X} \textbf{Y}}] $.  Compute $n_{\textbf{Y}}$ similarly.  Then the number of floating point numbers that must be transferred into the join is $p \times \left( n_{\textbf{X}} + n_{\textbf{Y}} \right)$, as each processor will receive one copy of a sub-tensor from the left, and from the right. Subsequently, we use $\textrm{cost}_\textrm{join}(\textbf{d}, \ell_\textbf{X}, \ell_\textbf{Y}, \textbf{b}_\textbf{XY})$ to refer to this quantity.

For example, consider the top-left case in Figure \ref{fig:dataflow}, where $\ell_\textbf{X} = [i, j]$, where $\ell_\textbf{Y} = [j, k]$, and $\ell_\textbf{XY} = [i, j, j, k]$. As $\textbf{b}_{\textbf{X} \textbf{Y}} = [8, 8, 8, 8]$ and $\textbf{d} = [4, 1, 1, 4],$ $\frac{\textbf{b}_{\textbf{X} \textbf{Y}}}{\textbf{d}} = [2, 8, 8, 2]$. Thus, $n_{\textbf{X}} = 2 \times 8 = 16$, $n_{\textbf{Y}} = 8 \times 2 = 16$, and thus $\textrm{cost}_\textrm{join}$ is $8 \times (16 + 16)$.

\vspace{5 pt}
\noindent
\textbf{Transferring into the aggregation.}  Let $n_{\textbf{Z}}$ be the number of floating point numbers in the sub-tensor resulting from each kernel call; this is $\prod \frac{\textbf{b}_{\textbf{X} \textbf{Y}}} {\textbf{d}} [ \ell_{\textbf{Z}} ; \ell_{\textbf{X} \textbf{Y}}] $.  Let $n_{\textrm{agg}}$ be the number of sub-tensors that are aggregated down to a single sub-tensor; this is $\prod {\textbf{d}} [ \ell_{\textrm{agg}} ; \ell_{\textbf{X} \textbf{Y}}] $.  The total number of floating point numbers that must be transferred is then 
$\frac{p}{n_{\textrm{agg}}} \left(n_{\textrm{agg}} - 1 \right) n_{\textbf{Z}}$.  There are $\frac{p}{n_{\textrm{agg}}}$ groups of tuples that must be aggregated.  In the best case, the amount of data transferred for each group is $\left(n_{\textrm{agg}} - 1 \right) n_{\textbf{Z}}$, as all tuples in a group are sent to a single processor for aggregation---but if a processor that already has such a tuple is chosen as the aggregation site, no transfer happens. We use $\textrm{cost}_\textrm{agg}(\textbf{d}, \ell_\textrm{agg}, \ell_\textbf{Z},  \ell_\textbf{XY}, \textbf{b}_\textbf{XY})$ to refer to this quantity.

For example, consider the top-left case in Figure 2. $\ell_{\textbf{agg}} = [j]$ so $n_{\textrm{agg}} = 1$, so the aggregation cost is zero (there is no aggregation).  Considering the bottom-right case where $\textbf{d} = [2, 2, 2, 4]$, we have $n_{\textrm{agg}} = 2$.
As
$\ell_{\textbf{Z}} = [i, k]$ we have $n_{\textbf{Z}} = 2 \times 4 = 8$, and the total number of floating point numbers moved is $\frac{16}{2} (2 - 1) 8 = 64$.

\begin{figure}[t]
\includegraphics[width=8.5cm]{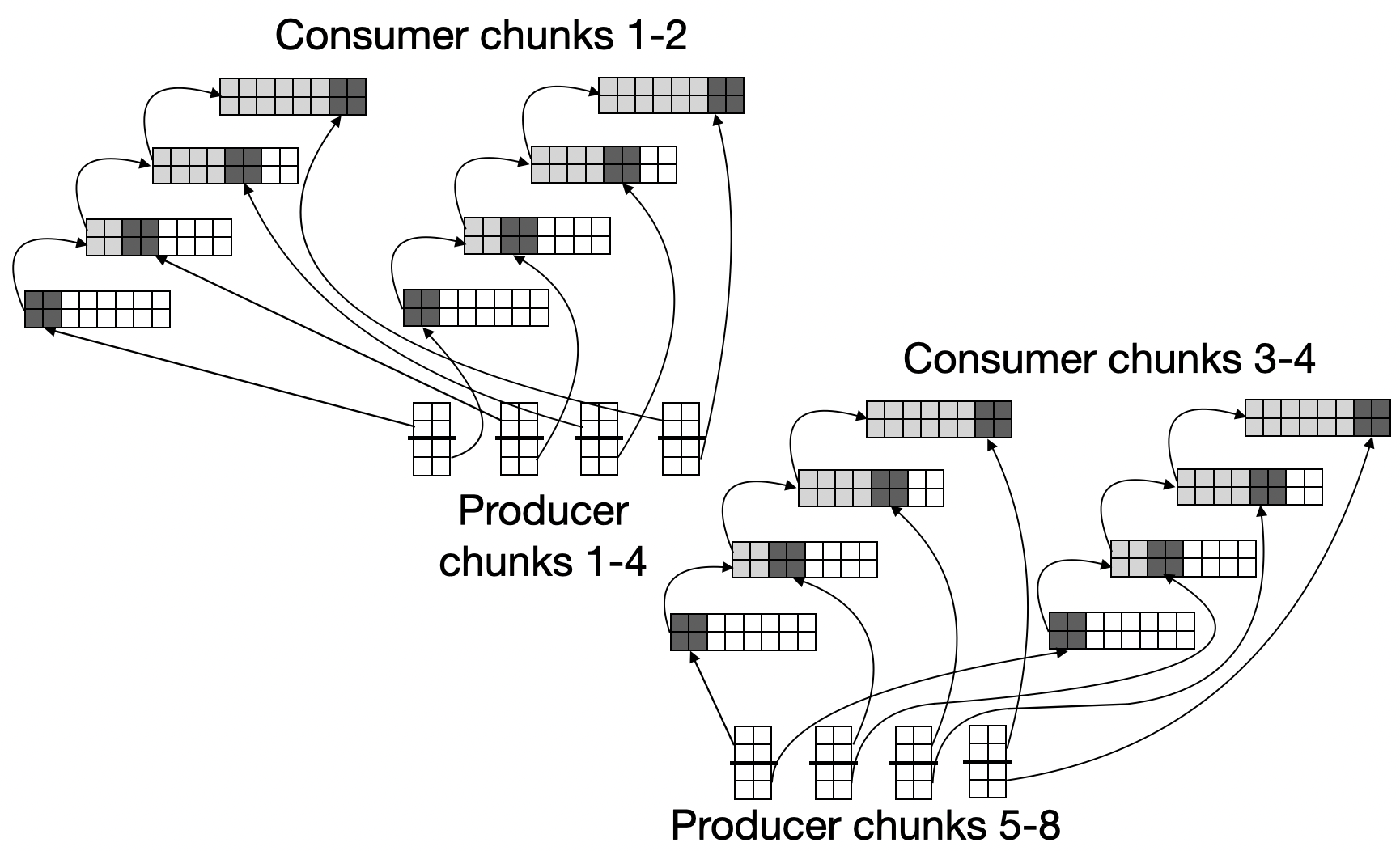}
\vspace{-15 pt}
 \caption{Modifying an \textsc{EinGraph} supplied by a programmer, to produce a \textsc{TaskGraph}, by adding bound vectors.  This is done so as to minimize an upper bound on the communication required for the corresponding decomposition.  }
 \vspace{-5 pt}
 \label{fig:schsys}
\end{figure}

\vspace{5 pt}
\noindent
\textbf{Re-partitioning across operations.} If two operations are connected (one is a producer and one is a consumer) but their partitionings do not match, re-partitioning is necessary. To understand how we cost a re-partition, imagine that we have two matrix multiplies described by the \texttt{EinSum} expression $\textbf{Z}_{i,k} \leftarrow \sum \textbf{X}_{i,j} \times \textbf{Y}_{j,k}$.  The partitioning of the first (the ``producer'') is described by $\textbf{d}^{(p)} = [2, 2, 2, 4]$ and the partitioning of the second (the ``consumer'') is described by $\textbf{d}^{(c)} = [4, 1, 1, 4]$.  Effectively, we are using the output of the computation in the lower left corner of Figure 2 as the left input to the computation in the upper right corner of Figure 2.

A graph showing the flow of data from the producer into the consumer is shown in Figure 4.  Each producer sub-tensor must be sent to two locations---one location where its top half is used, and one location where its bottom half is used.  As each producer sub-tensor has 8 floating point numbers, and there are 8 of them, the transfer cost is $2 \times 8 \times 8 = 128.$  Further, as we build up the consumer sub-tensors, each consumer sub-tensor must be sent to subsequent three locations after it accepts the first producer sub-tensor: to the location where it accepts the second, then the third, and the final.  Thus, the cost to move the input sub-tensors is $3 \times 16 \times 4 = 192$, for a total cost of 320.

Note that this computation can be optimized.  One could only send the half of the producer sub-tensor that is to be used at a location. However, as we want an upper bound on the cost, that is not dependent on implementation considerations, the cost is $128 + 192 = 320.$

In the general case, we have a producer producing a tensor with bound $\textbf{b}_\textbf{Z}$ and partitioning $\textbf{d}_\textbf{Z}$ and a consumer accepting a tensor with the same bound, and partitioning $\textbf{d}_\textbf{X}$.
Let $n_\textrm{p}$ be the number of floating point numbers in each producer sub-tensor, computed as $\prod \frac{\textbf{b}_{\textbf{Z}}} {\textbf{d}_\textbf{Z}}$.  This is $4 \times 2 = 8$ in our example. Let $n_\textrm{c}$ be the number of floating point numbers in each consumer sub-tensor; this is computed as $\prod \frac{\textbf{b}_{\textbf{X}}} {\textbf{d}_\textbf{Z}}$ ($2 \times 8 = 16$ in our example).  Let $n_\textrm{int}$ be the number of floating point numbers contributed by a producer sub-tensor to a consumer sub-tensor; this is $\prod \textrm{min} \left(\frac{\textbf{b}_{\textbf{Z}}} {\textbf{d}_\textbf{Z}}, 
\frac{\textbf{b}_{\textbf{X}}} {\textbf{d}_\textbf{Z}}
\right)$.  Here, $\textrm{min}()$ is computed element-wise. This is $2 \times 2 = 4$ in our example. And finally, 
let $n$ be the number of floating point numbers in the output of the producer, or the input to the consumer (computed as 
$n = \prod \textbf{d}_\textbf{Z}$; this is $8 \times 8 = 64$ in our example).  Then, the final cost is $\left( \frac{n_\textrm{c}}{n_\textrm{int}} - 1\right)\frac{n}{n_\textrm{c}}(n_\textrm{c} + n_\textrm{p})$, plus an additional transfer of $n_\textrm{p}\frac{n}{n_\textrm{c}}$ if $n_\textrm{p} \neq n_\textrm{int}$.  This latter term corresponds to the cost to send each producer sub-tensor to the location where part of it is extracted to form the initial consumer sub-tensor.  If the entire producer sub-tensor is used, then no such extraction is required.  Subsequently, we use $\textrm{cost}_\textrm{repart}(\textbf{d}_\textbf{X}, \textbf{d}_\textbf{Z}, \textbf{b}_\textbf{Z})$ to refer to this computation.

\section{The \texttt{EinDecomp} Algorithm}
\label{sec:Eindecomp}

Given an \textsc{EinGraph}, we consider how to label all of the bounds to produce a \textsc{TaskGraph} (as in Figure 3).

\subsection{Counting \texttt{EinSum} Partitionings}

The core reason there exists a tractable solution is that the number of partitionings to consider for a given \texttt{EinSum} expression can be kept surprisingly small.   
We first start by assuming that the number of processors $p = 2^N$ for integer $N$ and that every entry in $\textbf{d}$ will be chosen to be a power of two.  If the actual number of processors is not a power of two, $p$ can be chosen to be larger than the number of available processors. Choosing $p$ to be slightly larger than the number of processors physically available will increase the worst-case communication cost, but a quality assignment of operations to processors tends to alleviate this.  Thus, in our experience, forcing each $\textbf{d}$ to be a power of two has little effect on the upper bond that can be achieved.

If $D$ is the number of unique labels in $\ell_\textbf{X}$ and $\ell_\textbf{Y}$ then the number of possible values for $\textbf{d}$ is 
$\frac{(N + D - 1)!}{N! (D - 1)!}$.  This expression comes from the fact that we are effectively searching for all possible ways to place $N$ balls in $D$ buckets; placing a ``ball'' in a ``bucket'' doubles the value in the corresponding dimension of $\textbf{d}$.  Note that if two labels match across $\ell_\textbf{X}$ and $\ell_\textbf{Y}$ the corresponding dimensions are co-partitioned, and so they effectively count as one bucket; see Section 4.
This often allows all possible partitionings for an \texttt{EinSum} expression to be enumerated, brute-force. For example, if $N = 10$ and $D = 6$, then the number of partitionings is 3003.

\subsection{Dynamic Programming}

In an \textsc{EinGraph} where there is not more than one consumer for any non-input vertex, there exists a relatively efficient dynamic programming algorithm for computing an optimal \textsc{TaskGraph}.  This algorithm relies on a lookup table $M$ that is a map from (vertex $v$, partitioning $\textbf{d}_\textbf{Z}$) pairs to the lowest (optimal) cost for computing the subgraph up to and including vertex $v$, subject to the constraint that the output partition for the vertex is $\textbf{d}_\textbf{Z}$.

For example, imagine that we have an \textsc{EinGraph} vertex $v$ associated with matrix multiplication, which results in an output tensor with bound vector $[8, 8]$.  If we require $p =$ eight kernel calls in the implementation of $v$, the possible partitioning $\textbf{d}$ vectors associated with implementing the \texttt{EinSum} for $v$ are: $[2, 1, 1, 4];$ $[4, 1, 1, 2];$ $[8, 1, 1, 1];$ $[1, 1, 1, 8];$ $[2, 2, 2, 2];$ $[4, 2, 2, 1];$ $[1, 2, 2, 4];$ $[1, 8, 8, 1].$ Each of these partitioning vectors produces eight kernel calls since, after removing the repeated middle (join) index partitioning, the product over all entries is eight.  Thus, the set of possible output partitionings $\textbf{d}_\textbf{Z}$ for $v$ contains: $[2, 4];$ $[4, 2];$ $[8, 1];$ $[1, 8];$ $[2, 2];$ $[4, 1];$ $[1, 4];$ $[1, 1],$ as the middle dimensions are aggregated out in matrix multiplication.  The lookup table $M$ would then contain entries for $(v, [2, 4]),$ $(v, [4, 2]),$ $(v, [8, 1]),$ and so on. $M\left(v, [2, 4]\right),$ for example, would store the optimal cost for computing the  \textsc{EinGraph} up to vertex $v$, subject to the constraint that the \texttt{EinSum} expression associated with vertex $v$ produces an output partitioning of $\textbf{d}_\textbf{Z} = [2, 4]$.

\begin{figure}[t]
\includegraphics[width=8cm]{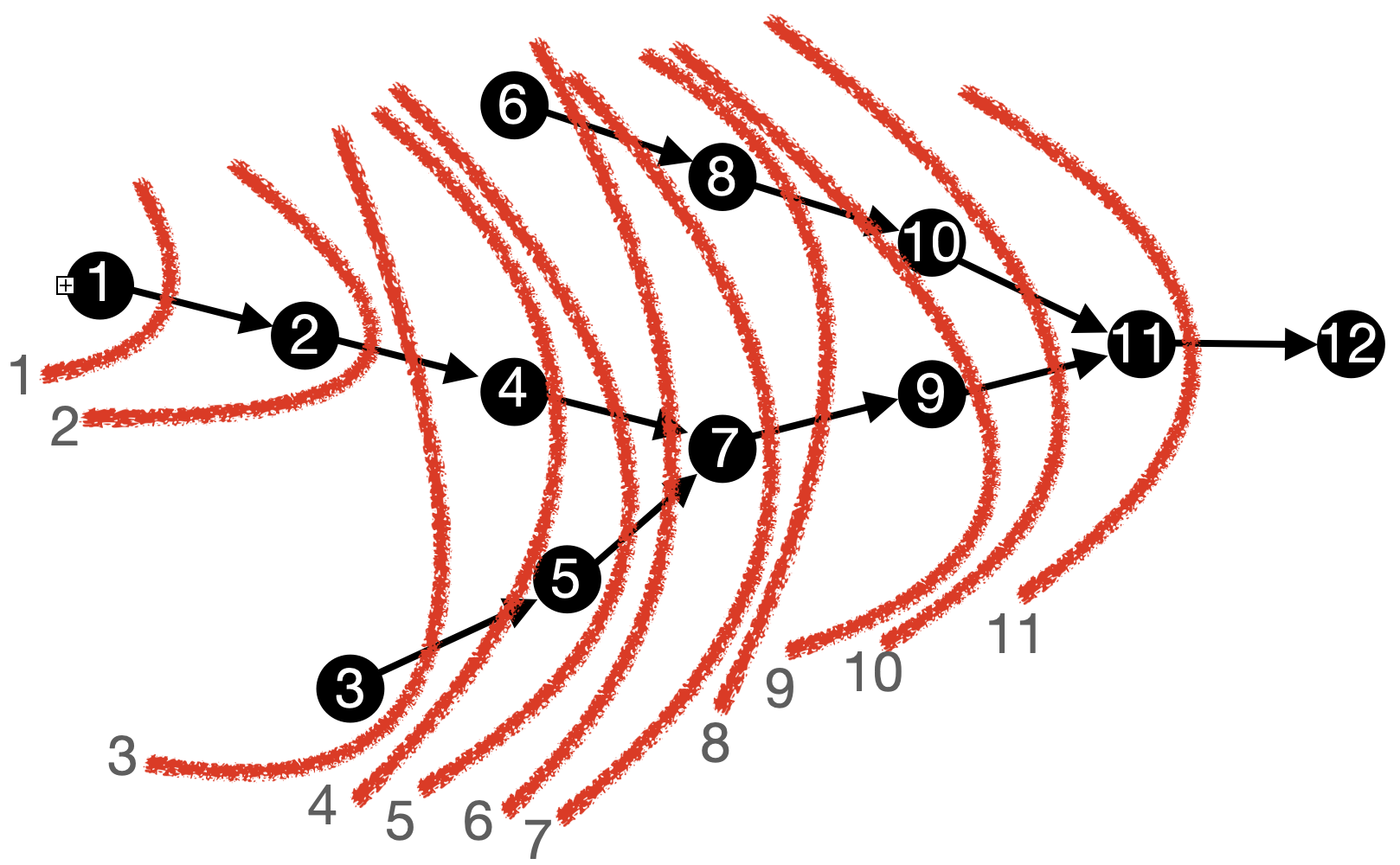}
 \vspace{-10 pt}
 \caption{Progression of the \texttt{EinDecmop} dynamic programming algorithm via a topological sort. After step 1, the lookup table $M$ holds lowest cost for producing all possible output partitionings of vertex 1. After step 2, $M$ holds the lowest costs for both vertex 1 and 2. And in general, after step $n$, $M$ holds the lowest cost for producing all possible output partitionings of vertices 1 through $n$.}
 \vspace{-5 pt}
 \label{fig:dp}
\end{figure}

The reason for maintaining this lookup table is that computing the lowest-cost implementation for vertex $v$ requires having access to the lowest cost implementations of the inputs to vertex $v$; if we have access to the lowest cost for every possible input partitioning to $v$, we can simply enumerate all of the partition vectors for $v$, applying the formulas of the previous section to figure out the best way to implement vertex $v$.  The overall dynamic programming algorithm proceeds according to the order provided by a topological sort of the input \textsc{EinGraph}, as shown in Figure \ref{fig:dp}. For any $v$ with no inputs, $M[v, \textbf{d}]$ to zero for each possible $\textbf{d}$.  This reflects the fact that inputs are generally pre-computed and pre-partitioned, offline, and incur no cost. 

\subsection{Computing the Optimal Cost During DP}

Our goal is to compute $M$ as the dynamic programming progresses. Consider a vertex $v$ with a binary \texttt{EinSum} expression for which we wish to compute $M[v, \textbf{d}_\textbf{Z}]$; let $v_\textbf{X}$ be the first entry in $v.\texttt{inputs}$ (so it corresponds to the $\textbf{X}$ input to the \texttt{EinSum} expression for $v$) and let let $v_\textbf{Y}$ be the second entry in $v.\texttt{inputs}$.  Let $\ell_\textbf{X}$, $\ell_\textbf{Y}$, and $\ell_\textbf{Z}$ to refer to the label vectors associated with $v.\texttt{EinSum}$, and $\textbf{b}_\textbf{XY}$ to refer to the bound vector for the \texttt{EinSum} computation.
Finally, let $\textrm{viable}(\texttt{EinSum}, p)$ return a list of all partitioning vectors for a tensor-relational implementation of the \texttt{EinSum} expression, subject to the constraint that the number of results of the embedded join is exactly $p$, ensuring $p$ pieces of parallel work.

\begin{figure}[t]
\includegraphics[width=8.5cm]{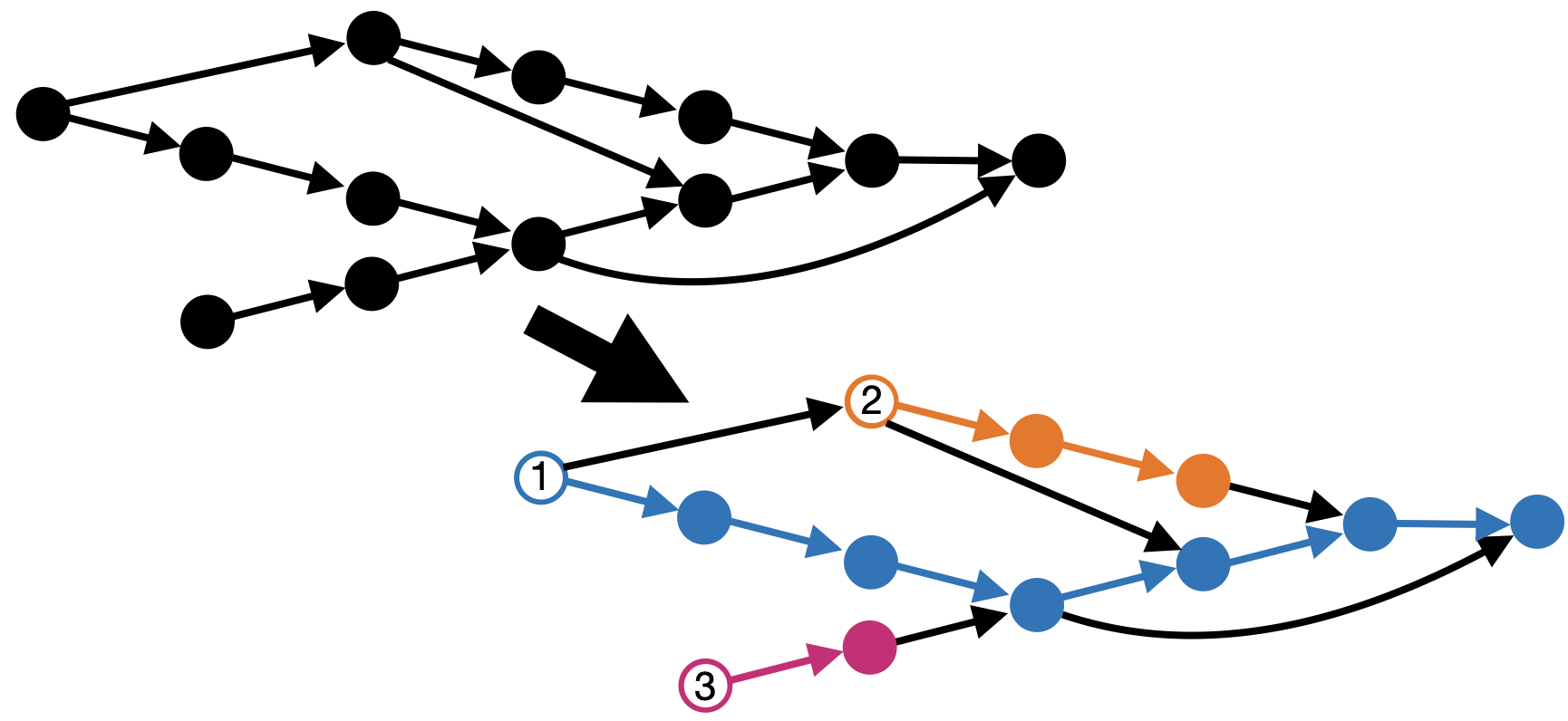}
\vspace{-15 pt}
 \caption{Linearizing an \textsc{EinGraph} to enable dynamic programming-based \textsc{TaskGraph} construction.   }
 \vspace{-10 pt}
 \label{fig:schsys}
\end{figure}

To compute $M[v, \textbf{d}_\textbf{Z}]$ we minimize:
\begin{align}
\textrm{over all } \textbf{d} \in \textrm{viable}(v.\texttt{EinSum}, p) \textrm{ where }\textbf{d}[\ell_\textbf{Z}; \ell_\textbf{XY}] = \textbf{d}_\textbf{Z}; \nonumber \\
\textrm{over all left input partitionings } \textbf{d}_{\textbf{X}}; \nonumber \\ 
\textrm{and over all right input partitionings } \textbf{d}_{\textbf{Y}}; \nonumber
\end{align}
the following expression:
\begin{align}
M[v_\textbf{X}, \textbf{d}_{\textbf{X}}] + 
M[v_\textbf{Y}, \textbf{d}_{\textbf{Y}}] + 
\textrm{cost}_\textrm{repart} (\textbf{d}_\textbf{X}, \textbf{d}[\ell_\textbf{Z}; \ell_\textbf{XY}], \textbf{b}_\textbf{XY}[\ell_\textbf{X}; \ell_\textbf{XY}]) + \nonumber \\
\textrm{cost}_\textrm{repart} (\textbf{d}_\textbf{Y}, \textbf{d}[\ell_\textbf{Z}; \ell_\textbf{XY}], \textbf{b}_\textbf{XY}[\ell_\textbf{Y}; \ell_\textbf{XY}]) + \nonumber \\
\textrm{cost}_\textrm{join}(\textbf{d}, \ell_\textbf{X}, \ell_\textbf{Y}, \textbf{b}_\textbf{XY}) + \textrm{cost}_\textrm{agg}(\textbf{d}, \ell_\textrm{agg}, \ell_\textbf{Z},  \ell_\textbf{XY}, \textbf{b}_\textbf{XY}) \nonumber
\end{align}

This is: we minimize the overall cost, which is the sum of the cost for computing the graph up to and including the left and right inputs, the cost to re-partition into the current \texttt{EinSum} expression, and then the cost to perform the join and aggregation for the current \texttt{EinSum} expression.  For a given output partitionings, this is minimized by considering all possible $\textbf{d}$ vectors that will produce that output; for each, we consider all possible left input partitionings, and all possible right input partitionings.

Once all $M[.]$ entries have been computed for the entire input \texttt{EinGraph}, the best labeling for the entire graph can be found by back-tracking from the best $M[v, \textbf{d}]$ value, where $v$ is the output vertex for the graph.  To produce the \textsc{TaskGraph}, at each vertex $v$ we choose the partitioning vector $\textbf{d}$ that produced the $M[v, \textbf{d}_\textbf{Z}]$ value that was used by the immediate descendent to produce its own optimal value.

\subsection{Handling General DAGs}

This algorithm is not applicable if there exists more than one consumer for the output of a non-input vertex, as the algorithm relies on being able to describe all possible optimal computations for a subgraph up to and including vertex $v$ with a set of $M[v, \textbf{d}]$ entries. If the output of $v$ has two consumers $v_1$ and $v_2$, we would need to maintain a lookup table that stores the optimal cost for all possible combinations of output partitionings for \emph{both} $v_1$ and $v_2$
Even if there is never more than one consumer of a non-input vertex, the algorithm may be too slow for a very large graph.

Such DAGs can be handled in approximate fashion by ``linearizing'' the graph. That is, we decompose the graph into a series of linear paths, and when optimizing, we consider only vertices and edges along the path.  For any inputs into the path that do not come from the path, we ignore the cost to compute the associated subgraph, as well as the possible re-partition cost.  So, for example, if the left ($v_\textbf{X}$) input into a vertex $v$ is along the path, but the right is not, 
to compute $M[v, \textbf{d}_\textbf{Z}]$ we  minimize:
\begin{align}
\textrm{over all } \textbf{d} \in \textrm{viable}(v.\texttt{EinSum}, p) \textrm{ where }\textbf{d}[\ell_\textbf{Z}; \ell_\textbf{XY}] = \textbf{d}_\textbf{Z}; \nonumber \\
\textrm{and over all left input partitionings } \textbf{d}_{\textbf{X}}; \nonumber 
\end{align}
the following expression:
\begin{align}
M[v_\textbf{X}, \textbf{d}_{\textbf{X}}] + 
\textrm{cost}_\textrm{repart} (\textbf{d}_\textbf{X}, \textbf{d}[\ell_\textbf{Z}; \ell_\textbf{XY}], \textbf{b}_\textbf{XY}[\ell_\textbf{X}; \ell_\textbf{XY}]) + \nonumber \\
\textrm{cost}_\textrm{join}(\textbf{d}, \ell_\textbf{X}, \ell_\textbf{Y}, \textbf{b}_\textbf{XY}) + \textrm{cost}_\textrm{agg}(\textbf{d}, \ell_\textrm{agg}, \ell_\textbf{Z},  \ell_\textbf{XY}, \textbf{b}_\textbf{XY}) \nonumber
\end{align}
The overall algorithm is shown above in Figure 5.  First, we find the longest path in the $\textsc{EinGraph}$ (shown in blue), that originates from vertex 1, and optimize only along that path.  To produce the \textbf{TaskGraph} partition labelings, if $v$ is the last vertex in the path, we choose the smallest $M[v, \textbf{d}_\textbf{Z}]$ value, and then backtrack back to vertex 1.  At each vertex $v$, we again choose the partitioning vector $\textbf{d}$ that produced the $M[v, \textbf{d}_\textbf{Z}]$ value that was used by the immediate descendent to produce its own optimal value.  Then, we find the next longest path (shown in orange, starting from vertex 2) and optimize along that path.  If $v$ is the last vertex along that path, we again choose the smallest $M[v, \textbf{d}_\textbf{Z}]$ value, and then backtrack back to vertex 2, labeling along the way.  This process is repeated for the third longest path, starting at vertex 3.

This algorithm does not find the globally optimal solution as it ignores re-partition costs \emph{across} the various paths.  In Figure 5, the algorithm will ignore the (possible) re-partition costs associated with the black edges in the graph, after linearization.  However, we have found little practical effect on the quality of the \textsc{TaskGraph} that is constructed, at least in the case of large language models.



\section{Experimental Evaluation}

\begin{figure}[t]
\includegraphics[width=8.7cm]{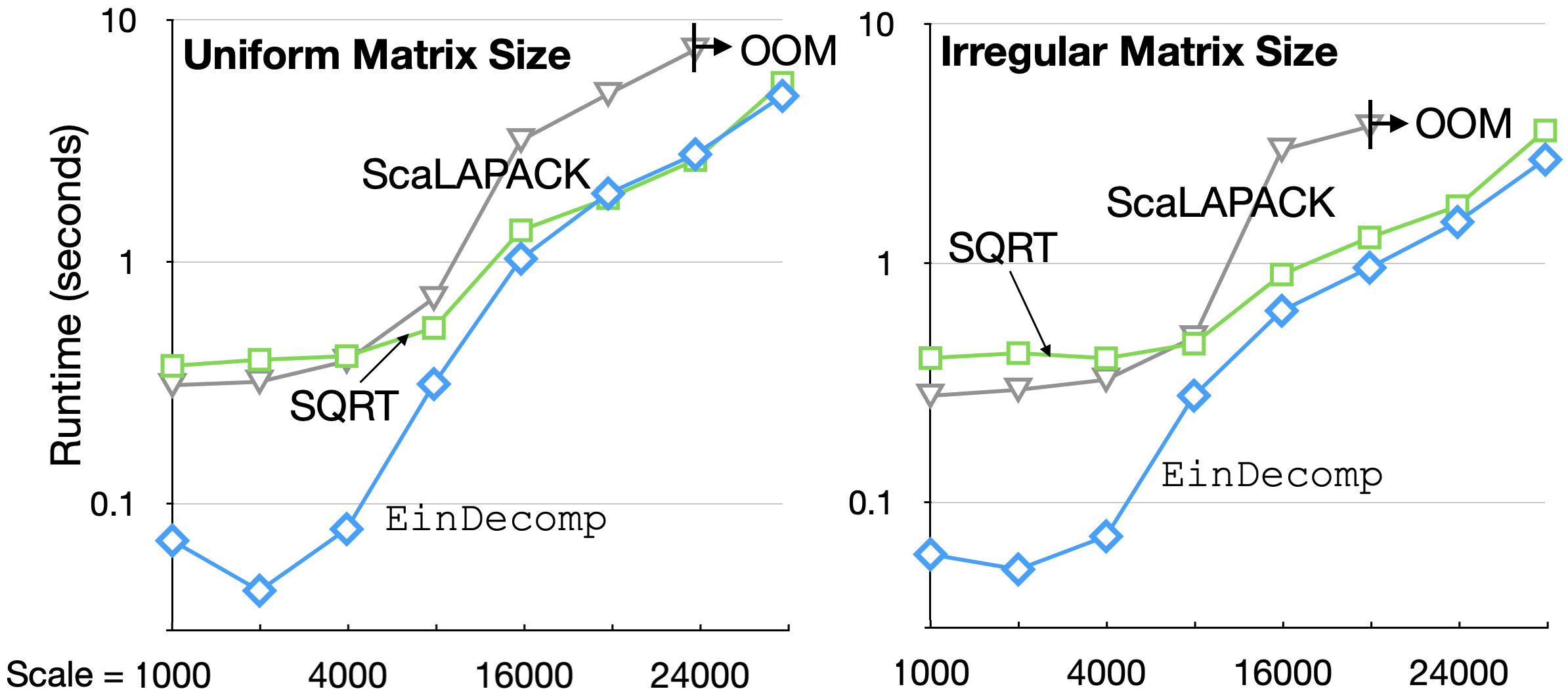}

 \caption{\texttt{EinDecmop} vs. SQRT vs. ScaLAPACK on a chain of matrix operations (CPU).}
 \vspace{-5 pt}
 \label{fig:Mat-Chain-CPU}
\end{figure}

\begin{figure}[t]
\includegraphics[width=8.7cm]{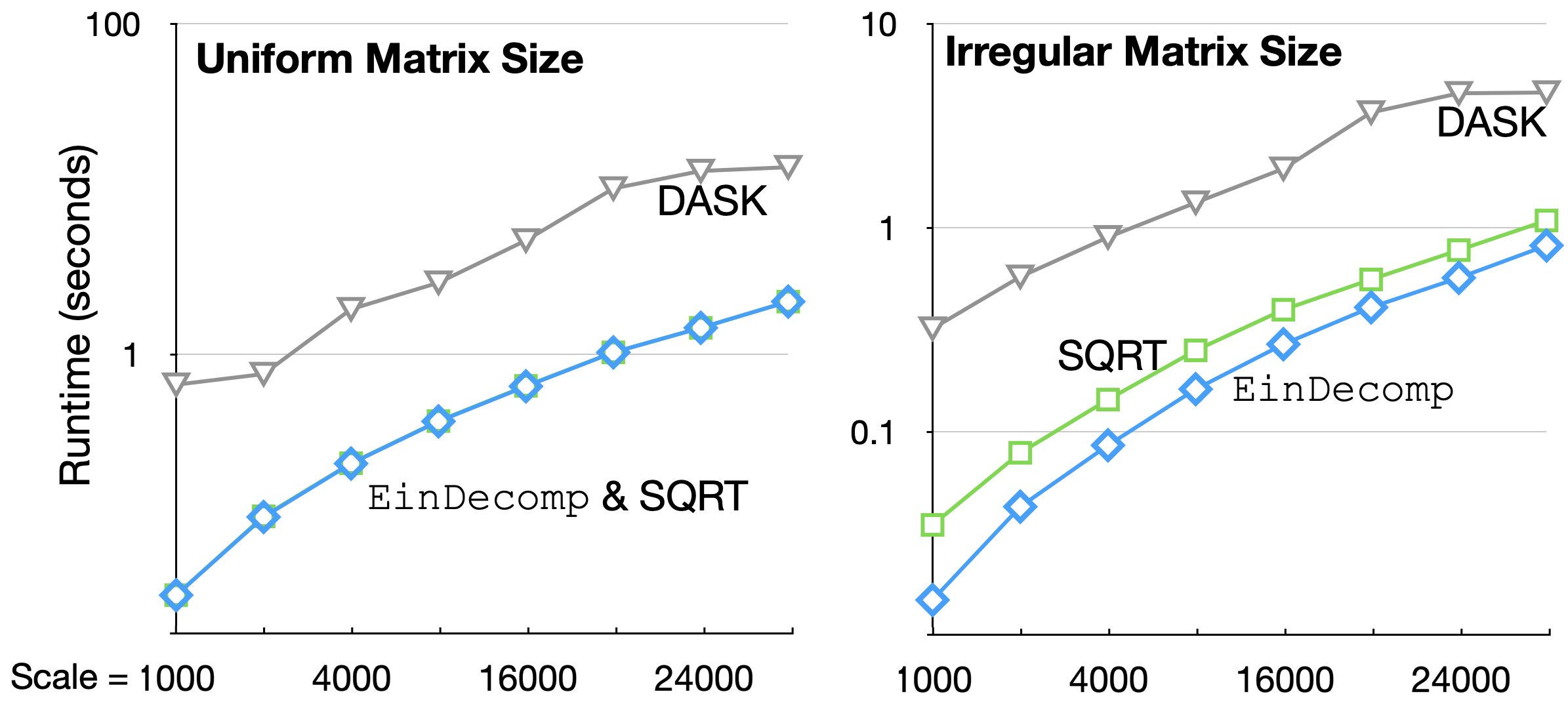}

 \caption{\texttt{EinDecmop} vs. SQRT vs. DASK on a chain of matrix operations (GPU).}
 \vspace{-5 pt}
 \label{fig:Mat-Chain-GPU}
\end{figure}

\begin{figure}[t]
\includegraphics[width=8.7cm]{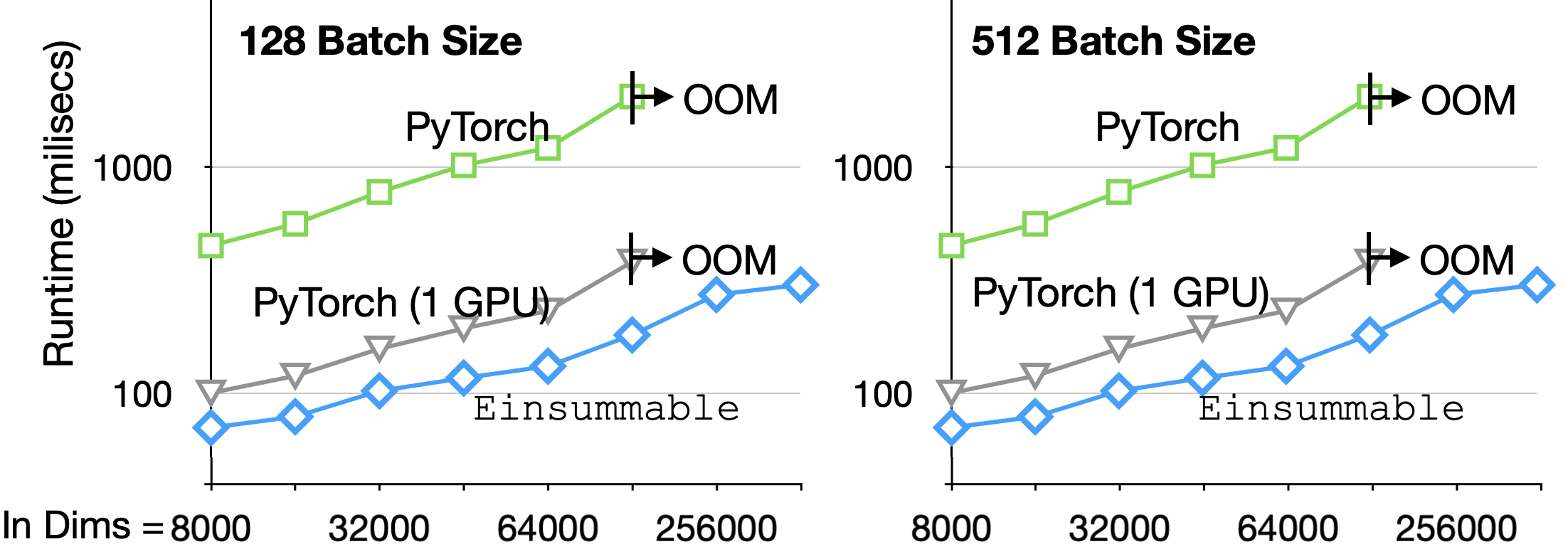}

 \caption{\texttt{EinDecmop} vs. PyTorch for training high-dimensional classifier.}
 \vspace{-5 pt}
 \label{fig:high-D}
\end{figure}

We experimentally evaluate the \texttt{EinDecomp} algorithm, with the goal of asking: how does the general \texttt{EinDecomp} approach compare to bespoke algorithms for decomposition?  For example, for large-scale matrix multiplication, how does \texttt{EinDecomp}  compare to the classical 3D algorithm \cite{agarwal1995three}?  For large-language model inference, how does \texttt{EinDecomp}  compare to the tensor-parallel approach taken by Megatron \cite{shoeybi2019megatron}?  A secondary question is: Can an \texttt{EinDecomp}-based system be competitive with other systems for large-scale machine learning?

\subsection{Experimental Overview}

We have implemented the \texttt{EinDecomp} algorithm on top of our \texttt{Einsummable} system, which is a machine learning system that utilizes an \texttt{EinSum}-based API and the \texttt{EinDecomp} decomposition algorithm to run machine learning computations on top of the \textsc{Turnip} execution engine \cite{ding2024turnip}.  \texttt{Einsummable} can run on both multi-machine CPU clusters, and on multi-GPU servers.  

\begin{figure*}[t]
\includegraphics[width=16cm]{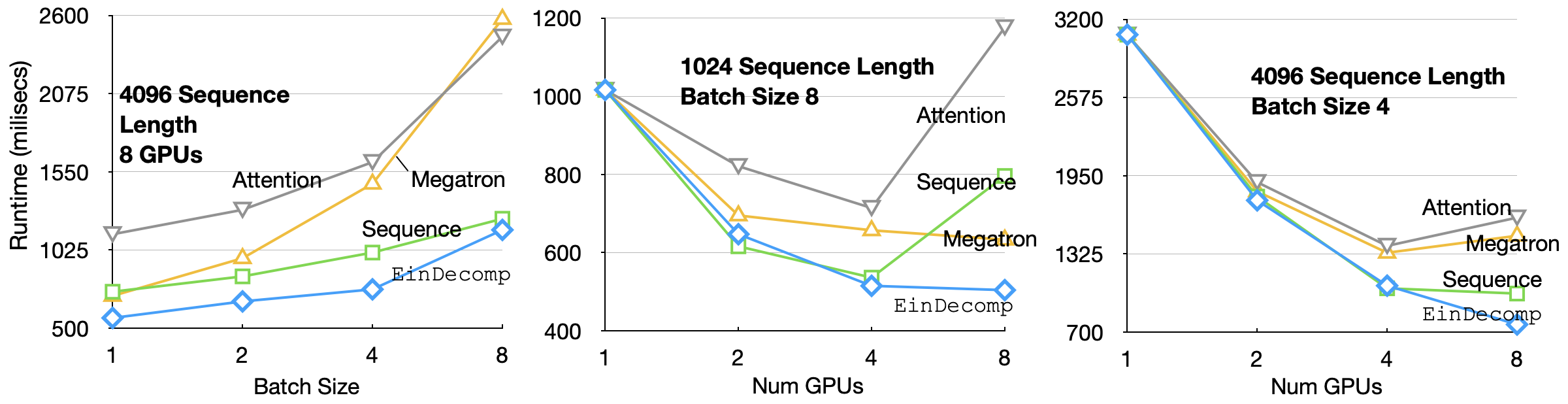}

 \caption{LLaMA large language model, run in \texttt{Einsummable}  using different decomposition algorithms.}
 \vspace{-5 pt}
 \label{fig:decomp-options}
\end{figure*}

In the case of CPU clusters, \texttt{EinSum} expressions are compiled by the system into kernels that (1) unpack the input tensors, (2) call Intel MKL's batch matrix multiply, and (3) re-pack the result into an output tensor. Since the packing and unpacking can be expensive and there are many ways to unpack tensors so that they are input into a batch matrix multiply \texttt{Einsummable} uses a cost-based optimizer to choose the best scheme.  Communication is implemented using the UCX \cite{shamis2015ucx} library.  In the case of GPU servers, \texttt{EinSum} expressions are compiled into GPU kernels using NVIDIA's CuTensor \cite{nvidia_cutensor}.
The \texttt{Einsummable} code base is currently around 27,000 lines (mostly C++).

CPU experiments  are run on Amazon Web Services (AWS), using a cluster of 16 \texttt{m6in.\-16xlarge} machines. Each machine has 256GB of RAM and 100 Gb per second network transfer. These machines have Intel Xeon processors (Ice Lake 8375C) and 32 physical cores, each of which has two virtual cores (used by Intel's hyper-threading implementation).  However, for the workloads we target (high-performance EinSum kernels), running one thread per physical core, pinned to the core, produces the best performance.   

GPU experiments are run on three different machines.  Some are run on AWS \texttt{P4d} instances with eight, 40 GB NVIDIA Tesla A100 GPUs.  The machine has two Intel CPUs with a total of 1.1TB of RAM.  Some are run in-house on server with four, 16GB NVIDIA Tesla P100 GPUs. This machine  has two Intel CPUs with a total of 1.3TB of RAM.  Some experiments are run on an in-house server with eight, 32GB NVIDIA Tesla P100 GPUs. This machine also has two Intel CPUs with a total of 1.3TB of RAM.

\subsection{Experiments Run}

Our evaluation of the \texttt{EinDecomp} framework considered in this paper consists of four different experiments.

\vspace{5 pt}
\noindent
\textbf{Experiment 1: } \textit{Testing} \texttt{EinDecomp}\textit{'s ability to parallelize large-scale matrix-chain arithmetic.}  Parallelizing chains of matrix operations is a classical problem, and it is a good test case because there are special-purpose high-performance computing softwares built specifically for this purpose.

We consider a chain of the form $(\textbf{A} \times \textbf{B}) + (\textbf{C} \times (\textbf{D} \times \textbf{E}))$.  There are two versions of the experiment.  In the first, all matrices are square, so for a scale of $s$, all matries are sized $s \times s$.  In the second, the matrices are sized as: \textbf{A}: $s \times .1s$,  \textbf{B}: $.1s \times s$, \textbf{C}: $s \times .1s$, \textbf{D}: $.1s \times 10s$, \textbf{E}: $10s \times s$.

Our experiments test a wide variety of $s$ values.  We run this on the CPU cluster and on the P100 GPU server.  On both, we compare \texttt{Einsummable} + \texttt{EinDecomp} with \texttt{Einsummable} + ``SQRT,'' where, to decompose a matrix into $n$ parts, we simply slice the matrix $\sqrt{n}$ ways vertically and  $\sqrt{n}$ ways horizontally.  If the matrices are square, this gives rise to the well-known 3D matrix multiplication algorithm, which is communication-optimal for square matrices.  On the CPU cluster we compare the two \texttt{Einsummable}-based algorithms with ScaLAPACK, which is the classical, high-performance, distributed matrix software.  On the GPU server we compare with DASK, which is a Python library for parallel computing.

CPU results are given in Figure  \ref{fig:Mat-Chain-CPU}.  GPU results are given in Figure  \ref{fig:Mat-Chain-GPU}.  ``OOM'' means that the method (ScaLAPACK in this case) failed due to out-of-memory errors.

\vspace{5 pt}
\noindent
\textbf{Experiment 2: } \textit{Testing} \texttt{EinDecomp}\textit{'s ability to decompose a large feed-forward neural network classifier for training.}  Here we consider training a large, feed-forward neural network (FFNN), using Pytorch (vanilla data parallel) and using \texttt{Einsummable} + \texttt{EinDecomp}.  We use the AmazonCat-14K	data set, which has 14,588 labels and 597,540 input features to train a FFNN having 8192 hidden neurons, using gradient descent.  We start with 8192 input features, and gradually increase the number of input features until all are used.  For PyTorch, we also test an option that uses only a single GPU, as opposed to all four P100 GPUs. Results are given in Figure  \ref{fig:high-D}, with batch sizes of 128 and 512 data points. 

\vspace{5 pt}
\noindent
\textbf{Experiment 3: } \textit{Comparing} \texttt{EinDecomp} \textit{with standard algorithms for decomposing a large language models.}
Our experiments target ``first token'' inference (``FTinf'') using LLaMA large-language model (or ``LLM''; FTinf is also known as ``prefill''): How long does it take to produce the first output token, given an input prompt?  LLaMA is MetaAI's signature large language model \cite{touvron2023llama}. FTinf is exceedingly expensive in terms of the amount of memory and compute required---both scale quadratically with the size of the prompt.  These experiments are run on the V100 server with eight GPUs.

Parallelizing LLM inference is very difficult, and there are a number of now-classical methods to do this, or at least, methods that may make sense in practice.  One alternative is the now-classic ``Megatron'' parallelization scheme \cite{shoeybi2019megatron}, which is a tensor-parallel or model-parallel scheme.  Another alternative is what we call ``sequence'', which splits the input sequence up $n$ ways, for inference on $n$ GPUs.  A final alternative is what we call ``attention'', where all attention heads are split into groups.  Of course, each of these methods is challenging to implement in a real-life LLM, and requires a lot of intricate decisions by the implementer.  To ensure an apples-to-apples comparison, all three of these methods were implemented on top of \texttt{Einsummable}, and compared with \texttt{Einsummable} + \texttt{EinDecomp} to see whether the automatic decomposition can compete with (or surpass) these other options.

We run three different FTinf experiments using the 7 billion parameter LLaMA model.  In the first, we use all eight GPUs, and perform FTinf on sequences of 4096 tokens, varying the batch size.  In the second, we use sequences of 1024 tokens, batch size eight, and vary the number of GPUs.  In the third, we uses sequences of 4096 tokens, batch size four, and vary the number of GPUs.  Results are given in Figure  \ref{fig:decomp-options}.

\begin{figure}[t]
\includegraphics[width=8.7cm]{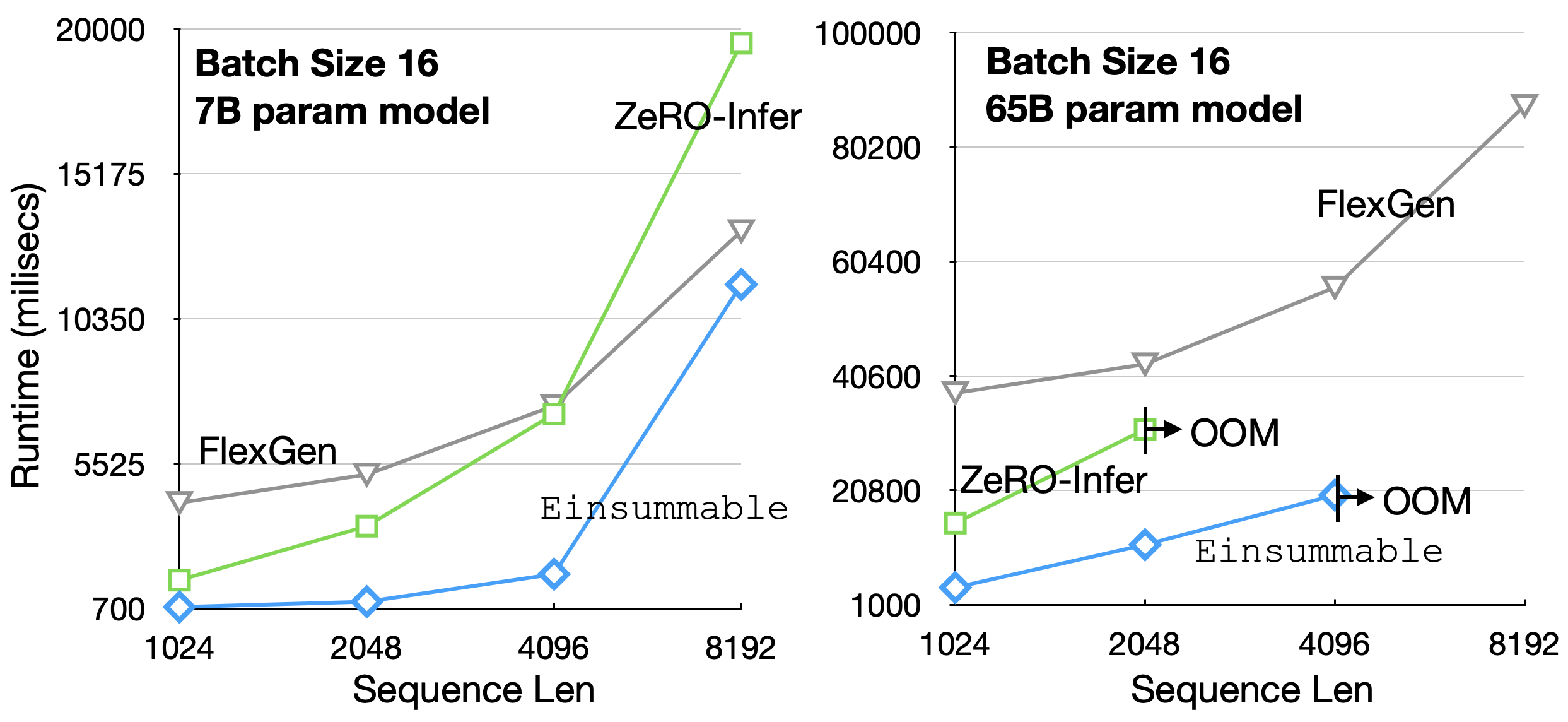}

 \caption{\texttt{Einsummable} vs ZeRO and FlexGen for LLaMA inference.}
 \vspace{-5 pt}
 \label{fig:ES-vs-others}
\end{figure}

\vspace{5 pt}
\noindent
\textbf{Experiment 4: } \textit{Comparing how an} \texttt{EinDecomp}\textit{-powered system compares with other systems for large-scale LLM inference.}  Our last experiment is tasked with asking the question: can an \texttt{EinDecomp}-based system compete with a standard, hand-coded system for LLM inference?  One unique aspect of \texttt{Einsummable} is that, due to the \textsc{Turnip} engine, it is able to page data in GPU RAM out to CPU RAM, and avoid the ubiquitous OOM errors that plague GPU computing.  Two other, well-known, PyTorch-based systems that can also do this are ZeRO Inference \cite{aminabadi2022deepspeed} and FlexGen \cite{sheng2023flexgen}.

We run two FTinf experiments on the AWS A100 server, both with a batch size of 16.  The first uses the 7 billion parameter LLaMA model, the second the 65 billion parameter model.  In each, we vary the sequence length.
Results are given in  Figure \ref{fig:ES-vs-others}.

\subsection{Discussion}

Over all experiments, with just a few exceptions, \texttt{Einsummable} + \texttt{Ein\-Decomp} was the best option tested.  Crucially, the automated parallelism enabled by \texttt{EinDecomp} almost always met or outperformed the performance of bespoke decomposition/parallelization strategies, such as the Megatron model parallelism or data parallelism.

\vspace{5 pt}
\noindent
In \textbf{Experiment 1}, it was quite surprising how poorly both ScaLAPACK and DASK performed.  In the GPU experiments, we observe exactly what as expected: \texttt{Einsum\-mable} + \texttt{EinDecomp} and \texttt{Einsum\-mable} + SQRT perform the same for uniform sizes, whereas there is a consistent $2\times$ gap for non-uniform sizes, as the simple decomposition provided by SQRT does not adapt to the skewed matrix sizes.  What is more  interesting is the GAP between \texttt{EinDecomp} and SQRT in the CPU experiments, even in the uniform case.  We suspect this is because the input matrices can be pre-placed, before execution, and this does not count against the overall time.  The \texttt{EinDecomp} may be able to exploit this. 

\vspace{5 pt}
\noindent
In \textbf{Experiment 2}, \texttt{EinDecomp} far outperformed data parallel PyTorch.  This is perhaps the worse case for a data parallel approach: a massive model which must be broadcast across GPUs, and a comparatively small input batch.  This is why PyTorch does so poorly.  In fact, PyTorch on one GPU did far better than PyTorch on four GPUs, because it removes the need to broadcast the model.  In comparison, \texttt{EinDecomp}  is able to automatically choose a much better plan than data parallel.

\vspace{5 pt}
\noindent
In \textbf{Experiment 3}, we find that in the case of powering a LLM, \texttt{EinDecomp} is able to consistently do as good as, or better than, all of the obvious alternatives for decomposing the model.  One surprising finding is how well decomposing along the sequence dimension works.   It is outperformed by \texttt{EinDecomp}, but consistently outperforms Megatron. 

\vspace{5 pt}
\noindent
Finally, in \textbf{Experiment 4}, we find that \texttt{Einsummable} + \texttt{EinDecomp} is able to far outperform other PyTorch-based systems that enable inference in a memory-constrained environment.  This does not directly evaluate \texttt{EinDecomp}, as the engine underlying \texttt{Einsummable} is very different from PyTorch.  However, it does show that the automated decompositions enabled by \texttt{EinDecomp} can power a high-performance machine learning system.  The parallelism in both ZeRO and FlexGen is carefully designed by hand.  For example, ZeRO uses a variant of data paralellism where the model is broadcast as needed, as inference works through the layers of the model. Yet, \texttt{Einsummable} + \texttt{EinDecomp} is able to do better than the bespoke ZeRO approach.

\section{Related Work}

The central idea in this paper, leveraging a fully declarative programming language to produce optimized decompositions of complex, ML and numerical computations, appears to be unique. However, variants of Einstein summation notation have been used for a long time (not surprisingly, Einstein used such notation \cite{einstein1938gravitational}), and it is supported by both TensorFlow and PyTorch.
Recent work has explored its use for ML \cite{laue2020simple} and its translation to SQL \cite{blacher2023efficient, pmlr-v202-tang23a}, though this sort of ``pure'' relational implementation cannot compete with a tensor-relational implementation in terms of performance.  There is a high computational overhead to push each tuple through a relational system \cite{neumann2011efficiently,sompolski2011vectorization,luo2018scalable}. The carefully-designed memory access patterns in an array-based kernel means that a CPU or GPU operates at close to its optimal floating-point operation per second rate, with little overhead.  

Parallelizing ML computations has generated a lot of recent interest.  Only a few recent works such as 
Alpa \cite{zheng2022alpa}, TensorOpt \cite{cai2021tensoropt}, FlexFlow \cite{jia2019beyond}, and Galvatron \cite{miao2022galvatron} have explored automatic parallelism, and those have considered the problem with varying levels of generality (Galvatron is relevant mostly for transformers, for example). None of these papers considered the question of how to declaratively specify the computation, and how to leverage that specification to facilitate auto-parallelism. Our use of Einstein summation notation is unique. 

Other efforts at supporting parallelism offer less automation, or are even more focused.  One of the most influential works in this domain is Megatron-LM \cite{shoeybi2019megatron}, but it proposes a specific parallelization scheme for transformers.   Amazon SageMaker Model Parallelism \cite{karakus2021amazon} is a PyTorch-based library that is designed to make training of large and complex models easier as is Microsoft's ZeRO-Infinity \cite{rajbhandari2021zero}.
GSPMD \cite{xu2021gspmd} is a  Google-built tool for distributed/parallel machine learning, built on top of Google's XLA compiler \cite{abadi2016tensorflow}.
Another effort is AutoMap \cite{schaarschmidt2021automap} from DeepMind.  AutoMap implements an operator-based abstraction, but programmers are asked to implement operators in a language (MLIR/PartIR \cite{lattner2021mlir,bondhugula2020high}) that forces users to expose parallelism. Mesh TensorFlow \cite{shazeer2018mesh} allows programmers to partition tensors across a compute mesh, but partitioning decisions are left to the programmer.  Other recent efforts (such as PyTorch Distributed \cite{li2020pytorch}) stick to the classical data parallel paradigm. 

This paper leverages the close connection between computations over tensors, and classical relational computations.  Others have noticed this connection before. The Tensor Relational Algebra, which we use in this paper, was proposed previously \cite{yuan2021tensor}. Many other systems leveraging this synergy have appeared in recent years, such as
SystemDS \cite{BoehmADGIKLPR20},
DAPHNE \cite{DammeBB0BCDDEFG22},
the Tensor Data Platform \cite{abs-2211-02753},
TileDB \cite{PapadopoulosDMM16},
ArrayQL \cite{SchuleGK022},
STOREL \cite{schleich2023optimizing}, and TensorDB \cite{kim2014tensordb,kim2014efficient}. 

\section{Conclusions}

We have described how to compile a computation consisting of a graph of operations expressed in the Einstein summation notation (\texttt{EinSum} language) and automatically decompose the \texttt{EinSum} operations to execute in a distributed CPU cluster, or on a GPU server.  We showed, through an extensive set of experiments, that the resulting  \texttt{EinDecomp} algorithm is, when implemented within the \texttt{Einsummable} system, able to perform as well as, or better than, many other standard parallelization options.

\balance
\bibliographystyle{ACM-Reference-Format}
\bibliography{SIGMOD2025a,SIGMOD2025b,SIGMOD2025c}


\begin{thebibliography}{57}


\ifx \showCODEN    \undefined \def \showCODEN     #1{\unskip}     \fi
\ifx \showDOI      \undefined \def \showDOI       #1{#1}\fi
\ifx \showISBNx    \undefined \def \showISBNx     #1{\unskip}     \fi
\ifx \showISBNxiii \undefined \def \showISBNxiii  #1{\unskip}     \fi
\ifx \showISSN     \undefined \def \showISSN      #1{\unskip}     \fi
\ifx \showLCCN     \undefined \def \showLCCN      #1{\unskip}     \fi
\ifx \shownote     \undefined \def \shownote      #1{#1}          \fi
\ifx \showarticletitle \undefined \def \showarticletitle #1{#1}   \fi
\ifx \showURL      \undefined \def \showURL       {\relax}        \fi
\providecommand\bibfield[2]{#2}
\providecommand\bibinfo[2]{#2}
\providecommand\natexlab[1]{#1}
\providecommand\showeprint[2][]{arXiv:#2}

\bibitem[Abadi et~al\mbox{.}(2016a)]%
        {abadi2016tensorflow}
\bibfield{author}{\bibinfo{person}{Mart{\'\i}n Abadi}, \bibinfo{person}{Paul Barham}, \bibinfo{person}{Jianmin Chen}, \bibinfo{person}{Zhifeng Chen}, \bibinfo{person}{Andy Davis}, \bibinfo{person}{Jeffrey Dean}, \bibinfo{person}{Matthieu Devin}, \bibinfo{person}{Sanjay Ghemawat}, \bibinfo{person}{Geoffrey Irving}, \bibinfo{person}{Michael Isard}, {et~al\mbox{.}}} \bibinfo{year}{2016}\natexlab{a}.
\newblock \showarticletitle{Tensorflow: A system for large-scale machine learning}. In \bibinfo{booktitle}{\emph{12th $\{$USENIX$\}$ symposium on operating systems design and implementation ($\{$OSDI$\}$ 16)}}. \bibinfo{pages}{265--283}.
\newblock


\bibitem[Abadi et~al\mbox{.}(2016b)]%
        {AbadiBCCDDDGIIK16}
\bibfield{author}{\bibinfo{person}{Mart{\'{\i}}n Abadi}, \bibinfo{person}{Paul Barham}, \bibinfo{person}{Jianmin Chen}, \bibinfo{person}{Zhifeng Chen}, \bibinfo{person}{Andy Davis}, \bibinfo{person}{Jeffrey Dean}, \bibinfo{person}{Matthieu Devin}, \bibinfo{person}{Sanjay Ghemawat}, \bibinfo{person}{Geoffrey Irving}, \bibinfo{person}{Michael Isard}, \bibinfo{person}{Manjunath Kudlur}, \bibinfo{person}{Josh Levenberg}, \bibinfo{person}{Rajat Monga}, \bibinfo{person}{Sherry Moore}, \bibinfo{person}{Derek~Gordon Murray}, \bibinfo{person}{Benoit Steiner}, \bibinfo{person}{Paul~A. Tucker}, \bibinfo{person}{Vijay Vasudevan}, \bibinfo{person}{Pete Warden}, \bibinfo{person}{Martin Wicke}, \bibinfo{person}{Yuan Yu}, {and} \bibinfo{person}{Xiaoqiang Zheng}.} \bibinfo{year}{2016}\natexlab{b}.
\newblock \showarticletitle{{TensorFlow: A System for Large-Scale Machine Learning}}. In \bibinfo{booktitle}{\emph{{OSDI}}}. \bibinfo{pages}{265--283}.
\newblock


\bibitem[Abdelfattah et~al\mbox{.}(2020)]%
        {abdelfattah2020matrix}
\bibfield{author}{\bibinfo{person}{Ahmad Abdelfattah}, \bibinfo{person}{Stanimire Tomov}, {and} \bibinfo{person}{Jack Dongarra}.} \bibinfo{year}{2020}\natexlab{}.
\newblock \showarticletitle{Matrix multiplication on batches of small matrices in half and half-complex precisions}.
\newblock \bibinfo{journal}{\emph{J. Parallel and Distrib. Comput.}}  \bibinfo{volume}{145} (\bibinfo{year}{2020}), \bibinfo{pages}{188--201}.
\newblock


\bibitem[Agarwal et~al\mbox{.}(1995)]%
        {agarwal1995three}
\bibfield{author}{\bibinfo{person}{Ramesh~C Agarwal}, \bibinfo{person}{Susanne~M Balle}, \bibinfo{person}{Fred~G Gustavson}, \bibinfo{person}{Mahesh Joshi}, {and} \bibinfo{person}{Prasad Palkar}.} \bibinfo{year}{1995}\natexlab{}.
\newblock \showarticletitle{A three-dimensional approach to parallel matrix multiplication}.
\newblock \bibinfo{journal}{\emph{IBM Journal of Research and Development}} \bibinfo{volume}{39}, \bibinfo{number}{5} (\bibinfo{year}{1995}), \bibinfo{pages}{575--582}.
\newblock


\bibitem[Aminabadi et~al\mbox{.}(2022)]%
        {aminabadi2022deepspeed}
\bibfield{author}{\bibinfo{person}{Reza~Yazdani Aminabadi}, \bibinfo{person}{Samyam Rajbhandari}, \bibinfo{person}{Ammar~Ahmad Awan}, \bibinfo{person}{Cheng Li}, \bibinfo{person}{Du Li}, \bibinfo{person}{Elton Zheng}, \bibinfo{person}{Olatunji Ruwase}, \bibinfo{person}{Shaden Smith}, \bibinfo{person}{Minjia Zhang}, \bibinfo{person}{Jeff Rasley}, {et~al\mbox{.}}} \bibinfo{year}{2022}\natexlab{}.
\newblock \showarticletitle{DeepSpeed-inference: enabling efficient inference of transformer models at unprecedented scale}. In \bibinfo{booktitle}{\emph{SC22: International Conference for High Performance Computing, Networking, Storage and Analysis}}. IEEE, \bibinfo{pages}{1--15}.
\newblock


\bibitem[Blacher et~al\mbox{.}(2023)]%
        {blacher2023efficient}
\bibfield{author}{\bibinfo{person}{Mark Blacher}, \bibinfo{person}{Julien Klaus}, \bibinfo{person}{Christoph Staudt}, \bibinfo{person}{S{\"o}ren Laue}, \bibinfo{person}{Viktor Leis}, {and} \bibinfo{person}{Joachim Giesen}.} \bibinfo{year}{2023}\natexlab{}.
\newblock \showarticletitle{Efficient and Portable Einstein Summation in SQL}.
\newblock \bibinfo{journal}{\emph{Proceedings of the ACM on Management of Data}} \bibinfo{volume}{1}, \bibinfo{number}{2} (\bibinfo{year}{2023}), \bibinfo{pages}{1--19}.
\newblock


\bibitem[Boehm et~al\mbox{.}(2020)]%
        {BoehmADGIKLPR20}
\bibfield{author}{\bibinfo{person}{Matthias Boehm}, \bibinfo{person}{Iulian Antonov}, \bibinfo{person}{Sebastian Baunsgaard}, \bibinfo{person}{Mark Dokter}, \bibinfo{person}{Robert Ginth{\"{o}}r}, \bibinfo{person}{Kevin Innerebner}, \bibinfo{person}{Florijan Klezin}, \bibinfo{person}{Stefanie~N. Lindstaedt}, \bibinfo{person}{Arnab Phani}, \bibinfo{person}{Benjamin Rath}, \bibinfo{person}{Berthold Reinwald}, \bibinfo{person}{Shafaq Siddiqui}, {and} \bibinfo{person}{Sebastian~Benjamin Wrede}.} \bibinfo{year}{2020}\natexlab{}.
\newblock \showarticletitle{{SystemDS: A Declarative Machine Learning System for the End-to-End Data Science Lifecycle}}. In \bibinfo{booktitle}{\emph{{CIDR}}}.
\newblock


\bibitem[Bondhugula(2020)]%
        {bondhugula2020high}
\bibfield{author}{\bibinfo{person}{Uday Bondhugula}.} \bibinfo{year}{2020}\natexlab{}.
\newblock \showarticletitle{High performance code generation in MLIR: An early case study with GEMM}.
\newblock \bibinfo{journal}{\emph{arXiv preprint arXiv:2003.00532}} (\bibinfo{year}{2020}).
\newblock


\bibitem[Cai et~al\mbox{.}(2021)]%
        {cai2021tensoropt}
\bibfield{author}{\bibinfo{person}{Zhenkun Cai}, \bibinfo{person}{Xiao Yan}, \bibinfo{person}{Kaihao Ma}, \bibinfo{person}{Yidi Wu}, \bibinfo{person}{Yuzhen Huang}, \bibinfo{person}{James Cheng}, \bibinfo{person}{Teng Su}, {and} \bibinfo{person}{Fan Yu}.} \bibinfo{year}{2021}\natexlab{}.
\newblock \showarticletitle{Tensoropt: Exploring the tradeoffs in distributed dnn training with auto-parallelism}.
\newblock \bibinfo{journal}{\emph{IEEE Transactions on Parallel and Distributed Systems}} \bibinfo{volume}{33}, \bibinfo{number}{8} (\bibinfo{year}{2021}), \bibinfo{pages}{1967--1981}.
\newblock


\bibitem[Chen et~al\mbox{.}(2018)]%
        {chen2018tvm}
\bibfield{author}{\bibinfo{person}{Tianqi Chen}, \bibinfo{person}{Thierry Moreau}, \bibinfo{person}{Ziheng Jiang}, \bibinfo{person}{Lianmin Zheng}, \bibinfo{person}{Eddie Yan}, \bibinfo{person}{Haichen Shen}, \bibinfo{person}{Meghan Cowan}, \bibinfo{person}{Leyuan Wang}, \bibinfo{person}{Yuwei Hu}, \bibinfo{person}{Luis Ceze}, {et~al\mbox{.}}} \bibinfo{year}{2018}\natexlab{}.
\newblock \showarticletitle{$\{$TVM$\}$: An automated end-to-end optimizing compiler for deep learning}. In \bibinfo{booktitle}{\emph{13th $\{$USENIX$\}$ Symposium on Operating Systems Design and Implementation ($\{$OSDI$\}$ 18)}}. \bibinfo{pages}{578--594}.
\newblock


\bibitem[Choi et~al\mbox{.}(1992)]%
        {choi1992scalapack}
\bibfield{author}{\bibinfo{person}{Jaeyoung Choi}, \bibinfo{person}{Jack~J Dongarra}, \bibinfo{person}{Roldan Pozo}, {and} \bibinfo{person}{David~W Walker}.} \bibinfo{year}{1992}\natexlab{}.
\newblock \showarticletitle{ScaLAPACK: A scalable linear algebra library for distributed memory concurrent computers}. In \bibinfo{booktitle}{\emph{The Fourth Symposium on the Frontiers of Massively Parallel Computation}}. IEEE Computer Society, \bibinfo{pages}{120--121}.
\newblock


\bibitem[Corporation(2023)]%
        {nvidia_cutensor}
\bibfield{author}{\bibinfo{person}{Nvidia Corporation}.} \bibinfo{year}{2023}\natexlab{}.
\newblock \bibinfo{title}{cuTENSOR: A CUDA Library for Tensor Algebra}.
\newblock
\newblock
\urldef\tempurl%
\url{https://docs.nvidia.com/cuda/cutensor/latest/index.html}
\showURL{%
\tempurl}
\newblock
\shownote{Accessed: 2024-10-01}.


\bibitem[Damme et~al\mbox{.}(2022)]%
        {DammeBB0BCDDEFG22}
\bibfield{author}{\bibinfo{person}{Patrick Damme} {et~al\mbox{.}}} \bibinfo{year}{2022}\natexlab{}.
\newblock \showarticletitle{{DAPHNE:} An Open and Extensible System Infrastructure for Integrated Data Analysis Pipelines}. In \bibinfo{booktitle}{\emph{{CIDR}}}.
\newblock
\urldef\tempurl%
\url{https://www.cidrdb.org/cidr2022/papers/p4-damme.pdf}
\showURL{%
\tempurl}


\bibitem[Dean et~al\mbox{.}(2012)]%
        {dean2012large}
\bibfield{author}{\bibinfo{person}{Jeffrey Dean}, \bibinfo{person}{Greg Corrado}, \bibinfo{person}{Rajat Monga}, \bibinfo{person}{Kai Chen}, \bibinfo{person}{Matthieu Devin}, \bibinfo{person}{Mark Mao}, \bibinfo{person}{Marc'aurelio Ranzato}, \bibinfo{person}{Andrew Senior}, \bibinfo{person}{Paul Tucker}, \bibinfo{person}{Ke Yang}, {et~al\mbox{.}}} \bibinfo{year}{2012}\natexlab{}.
\newblock \showarticletitle{Large scale distributed deep networks}. In \bibinfo{booktitle}{\emph{Advances in neural information processing systems}}. \bibinfo{pages}{1223--1231}.
\newblock


\bibitem[Ding et~al\mbox{.}(2024)]%
        {ding2024turnip}
\bibfield{author}{\bibinfo{person}{Zhimin Ding}, \bibinfo{person}{Jiawen Yao}, \bibinfo{person}{Brianna Barrow}, \bibinfo{person}{Tania~Lorido Botran}, \bibinfo{person}{Christopher Jermaine}, \bibinfo{person}{Yuxin Tang}, \bibinfo{person}{Jiehui Li}, \bibinfo{person}{Xinyu Yao}, \bibinfo{person}{Sleem~Mahmoud Abdelghafar}, {and} \bibinfo{person}{Daniel Bourgeois}.} \bibinfo{year}{2024}\natexlab{}.
\newblock \showarticletitle{TURNIP: A" Nondeterministic" GPU Runtime with CPU RAM Offload}.
\newblock \bibinfo{journal}{\emph{arXiv preprint arXiv:2405.16283}} (\bibinfo{year}{2024}).
\newblock


\bibitem[Einstein et~al\mbox{.}(1938)]%
        {einstein1938gravitational}
\bibfield{author}{\bibinfo{person}{Albert Einstein}, \bibinfo{person}{Leopold Infeld}, {and} \bibinfo{person}{Banesh Hoffmann}.} \bibinfo{year}{1938}\natexlab{}.
\newblock \showarticletitle{The gravitational equations and the problem of motion}.
\newblock \bibinfo{journal}{\emph{Annals of mathematics}} (\bibinfo{year}{1938}), \bibinfo{pages}{65--100}.
\newblock


\bibitem[Farber and Asanovic(1997)]%
        {farber1997parallel}
\bibfield{author}{\bibinfo{person}{Philipp Farber} {and} \bibinfo{person}{Krste Asanovic}.} \bibinfo{year}{1997}\natexlab{}.
\newblock \showarticletitle{Parallel neural network training on multi-spert}. In \bibinfo{booktitle}{\emph{Algorithms and Architectures for Parallel Processing, 1997. ICAPP 97., 1997 3rd International Conference on}}. IEEE, \bibinfo{pages}{659--666}.
\newblock


\bibitem[Gandhi et~al\mbox{.}(2022)]%
        {abs-2211-02753}
\bibfield{author}{\bibinfo{person}{Apurva Gandhi}, \bibinfo{person}{Yuki Asada}, \bibinfo{person}{Victor Fu}, \bibinfo{person}{Advitya Gemawat}, \bibinfo{person}{Lihao Zhang}, \bibinfo{person}{Rathijit Sen}, \bibinfo{person}{Carlo Curino}, \bibinfo{person}{Jes{\'{u}}s Camacho{-}Rodr{\'{\i}}guez}, {and} \bibinfo{person}{Matteo Interlandi}.} \bibinfo{year}{2022}\natexlab{}.
\newblock \showarticletitle{The Tensor Data Platform: Towards an AI-centric Database System}.
\newblock \bibinfo{journal}{\emph{CoRR}}  \bibinfo{volume}{abs/2211.02753} (\bibinfo{year}{2022}).
\newblock
\urldef\tempurl%
\url{https://doi.org/10.48550/arXiv.2211.02753}
\showDOI{\tempurl}


\bibitem[Hadjis et~al\mbox{.}(2016)]%
        {hadjis2016omnivore}
\bibfield{author}{\bibinfo{person}{Stefan Hadjis}, \bibinfo{person}{Ce Zhang}, \bibinfo{person}{Ioannis Mitliagkas}, \bibinfo{person}{Dan Iter}, {and} \bibinfo{person}{Christopher R{\'e}}.} \bibinfo{year}{2016}\natexlab{}.
\newblock \showarticletitle{Omnivore: An optimizer for multi-device deep learning on {CPU}s and {GPU}s}.
\newblock \bibinfo{journal}{\emph{arXiv preprint arXiv:1606.04487}} (\bibinfo{year}{2016}).
\newblock


\bibitem[Hasan and Motwani(1994)]%
        {hasan1994optimization}
\bibfield{author}{\bibinfo{person}{Waqar Hasan} {and} \bibinfo{person}{Rajeev Motwani}.} \bibinfo{year}{1994}\natexlab{}.
\newblock \showarticletitle{Optimization algorithms for exploiting the parallelism-communication tradeoff in pipelined parallelism}. In \bibinfo{booktitle}{\emph{VLDB}}, Vol.~\bibinfo{volume}{94}. Citeseer, \bibinfo{pages}{12--15}.
\newblock


\bibitem[Huang et~al\mbox{.}(2019)]%
        {huang2019gpipe}
\bibfield{author}{\bibinfo{person}{Yanping Huang}, \bibinfo{person}{Youlong Cheng}, \bibinfo{person}{Ankur Bapna}, \bibinfo{person}{Orhan Firat}, \bibinfo{person}{Dehao Chen}, \bibinfo{person}{Mia Chen}, \bibinfo{person}{HyoukJoong Lee}, \bibinfo{person}{Jiquan Ngiam}, \bibinfo{person}{Quoc~V Le}, \bibinfo{person}{Yonghui Wu}, {et~al\mbox{.}}} \bibinfo{year}{2019}\natexlab{}.
\newblock \showarticletitle{Gpipe: Efficient training of giant neural networks using pipeline parallelism}. In \bibinfo{booktitle}{\emph{Advances in neural information processing systems}}. \bibinfo{pages}{103--112}.
\newblock


\bibitem[Jia et~al\mbox{.}(2019)]%
        {jia2019beyond}
\bibfield{author}{\bibinfo{person}{Zhihao Jia}, \bibinfo{person}{Matei Zaharia}, {and} \bibinfo{person}{Alex Aiken}.} \bibinfo{year}{2019}\natexlab{}.
\newblock \showarticletitle{Beyond Data and Model Parallelism for Deep Neural Networks.}
\newblock \bibinfo{journal}{\emph{Proceedings of Machine Learning and Systems}}  \bibinfo{volume}{1} (\bibinfo{year}{2019}), \bibinfo{pages}{1--13}.
\newblock


\bibitem[Karakus et~al\mbox{.}(2021)]%
        {karakus2021amazon}
\bibfield{author}{\bibinfo{person}{Can Karakus}, \bibinfo{person}{Rahul Huilgol}, \bibinfo{person}{Fei Wu}, \bibinfo{person}{Anirudh Subramanian}, \bibinfo{person}{Cade Daniel}, \bibinfo{person}{Derya Cavdar}, \bibinfo{person}{Teng Xu}, \bibinfo{person}{Haohan Chen}, \bibinfo{person}{Arash Rahnama}, {and} \bibinfo{person}{Luis Quintela}.} \bibinfo{year}{2021}\natexlab{}.
\newblock \showarticletitle{Amazon {SageMaker} Model Parallelism: A General and Flexible Framework for Large Model Training}.
\newblock \bibinfo{journal}{\emph{arXiv preprint arXiv:2111.05972}} (\bibinfo{year}{2021}).
\newblock


\bibitem[Kim and Candan(2014a)]%
        {kim2014efficient}
\bibfield{author}{\bibinfo{person}{Mijung Kim} {and} \bibinfo{person}{K~Sel{\c{c}}uk Candan}.} \bibinfo{year}{2014}\natexlab{a}.
\newblock \showarticletitle{Efficient static and dynamic in-database tensor decompositions on chunk-based array stores}. In \bibinfo{booktitle}{\emph{Proceedings of the 23rd ACM International Conference on Conference on Information and Knowledge Management}}. \bibinfo{pages}{969--978}.
\newblock


\bibitem[Kim and Candan(2014b)]%
        {kim2014tensordb}
\bibfield{author}{\bibinfo{person}{Mijung Kim} {and} \bibinfo{person}{K~Sel{\c{c}}uk Candan}.} \bibinfo{year}{2014}\natexlab{b}.
\newblock \showarticletitle{Tensordb: In-database tensor manipulation with tensor-relational query plans}. In \bibinfo{booktitle}{\emph{Proceedings of the 23rd ACM International Conference on Conference on Information and Knowledge Management}}. ACM, \bibinfo{pages}{2039--2041}.
\newblock


\bibitem[Kjolstad et~al\mbox{.}(2017)]%
        {kjolstad2017tensor}
\bibfield{author}{\bibinfo{person}{Fredrik Kjolstad}, \bibinfo{person}{Shoaib Kamil}, \bibinfo{person}{Stephen Chou}, \bibinfo{person}{David Lugato}, {and} \bibinfo{person}{Saman Amarasinghe}.} \bibinfo{year}{2017}\natexlab{}.
\newblock \showarticletitle{The tensor algebra compiler}.
\newblock \bibinfo{journal}{\emph{Proceedings of the ACM on Programming Languages}} \bibinfo{volume}{1}, \bibinfo{number}{OOPSLA} (\bibinfo{year}{2017}), \bibinfo{pages}{1--29}.
\newblock


\bibitem[Lattner et~al\mbox{.}(2021)]%
        {lattner2021mlir}
\bibfield{author}{\bibinfo{person}{Chris Lattner}, \bibinfo{person}{Mehdi Amini}, \bibinfo{person}{Uday Bondhugula}, \bibinfo{person}{Albert Cohen}, \bibinfo{person}{Andy Davis}, \bibinfo{person}{Jacques Pienaar}, \bibinfo{person}{River Riddle}, \bibinfo{person}{Tatiana Shpeisman}, \bibinfo{person}{Nicolas Vasilache}, {and} \bibinfo{person}{Oleksandr Zinenko}.} \bibinfo{year}{2021}\natexlab{}.
\newblock \showarticletitle{Mlir: Scaling compiler infrastructure for domain specific computation}. In \bibinfo{booktitle}{\emph{2021 IEEE/ACM International Symposium on Code Generation and Optimization (CGO)}}. IEEE, \bibinfo{pages}{2--14}.
\newblock


\bibitem[Laue et~al\mbox{.}(2020)]%
        {laue2020simple}
\bibfield{author}{\bibinfo{person}{S{\"o}ren Laue}, \bibinfo{person}{Matthias Mitterreiter}, {and} \bibinfo{person}{Joachim Giesen}.} \bibinfo{year}{2020}\natexlab{}.
\newblock \showarticletitle{A simple and efficient tensor calculus}. In \bibinfo{booktitle}{\emph{Proceedings of the AAAI Conference on Artificial Intelligence}}, Vol.~\bibinfo{volume}{34}. \bibinfo{pages}{4527--4534}.
\newblock


\bibitem[Li et~al\mbox{.}(2020)]%
        {li2020pytorch}
\bibfield{author}{\bibinfo{person}{Shen Li}, \bibinfo{person}{Yanli Zhao}, \bibinfo{person}{Rohan Varma}, \bibinfo{person}{Omkar Salpekar}, \bibinfo{person}{Pieter Noordhuis}, \bibinfo{person}{Teng Li}, \bibinfo{person}{Adam Paszke}, \bibinfo{person}{Jeff Smith}, \bibinfo{person}{Brian Vaughan}, \bibinfo{person}{Pritam Damania}, {et~al\mbox{.}}} \bibinfo{year}{2020}\natexlab{}.
\newblock \showarticletitle{PyTorch Distributed: Experiences on Accelerating Data Parallel Training}.
\newblock \bibinfo{journal}{\emph{arXiv preprint arXiv:2006.15704}} (\bibinfo{year}{2020}).
\newblock


\bibitem[Luo et~al\mbox{.}(2018)]%
        {luo2018scalable}
\bibfield{author}{\bibinfo{person}{Shangyu Luo}, \bibinfo{person}{Zekai~J Gao}, \bibinfo{person}{Michael Gubanov}, \bibinfo{person}{Luis~L Perez}, {and} \bibinfo{person}{Christopher Jermaine}.} \bibinfo{year}{2018}\natexlab{}.
\newblock \showarticletitle{Scalable linear algebra on a relational database system}.
\newblock \bibinfo{journal}{\emph{IEEE Transactions on Knowledge and Data Engineering}} \bibinfo{volume}{31}, \bibinfo{number}{7} (\bibinfo{year}{2018}), \bibinfo{pages}{1224--1238}.
\newblock


\bibitem[Mehta and DeWitt(1995)]%
        {mehta1995managing}
\bibfield{author}{\bibinfo{person}{Manish Mehta} {and} \bibinfo{person}{David~J DeWitt}.} \bibinfo{year}{1995}\natexlab{}.
\newblock \showarticletitle{Managing intra-operator parallelism in parallel database systems}. In \bibinfo{booktitle}{\emph{VLDB}}, Vol.~\bibinfo{volume}{95}. \bibinfo{pages}{382--394}.
\newblock


\bibitem[Miao et~al\mbox{.}(2022)]%
        {miao2022galvatron}
\bibfield{author}{\bibinfo{person}{Xupeng Miao}, \bibinfo{person}{Yujie Wang}, \bibinfo{person}{Youhe Jiang}, \bibinfo{person}{Chunan Shi}, \bibinfo{person}{Xiaonan Nie}, \bibinfo{person}{Hailin Zhang}, {and} \bibinfo{person}{Bin Cui}.} \bibinfo{year}{2022}\natexlab{}.
\newblock \showarticletitle{Galvatron: Efficient transformer training over multiple gpus using automatic parallelism}.
\newblock \bibinfo{journal}{\emph{arXiv preprint arXiv:2211.13878}} (\bibinfo{year}{2022}).
\newblock


\bibitem[Narayanan et~al\mbox{.}(2019)]%
        {narayanan2019pipedream}
\bibfield{author}{\bibinfo{person}{Deepak Narayanan}, \bibinfo{person}{Aaron Harlap}, \bibinfo{person}{Amar Phanishayee}, \bibinfo{person}{Vivek Seshadri}, \bibinfo{person}{Nikhil~R Devanur}, \bibinfo{person}{Gregory~R Ganger}, \bibinfo{person}{Phillip~B Gibbons}, {and} \bibinfo{person}{Matei Zaharia}.} \bibinfo{year}{2019}\natexlab{}.
\newblock \showarticletitle{PipeDream: generalized pipeline parallelism for DNN training}. In \bibinfo{booktitle}{\emph{Proceedings of the 27th ACM symposium on operating systems principles}}. \bibinfo{pages}{1--15}.
\newblock


\bibitem[Neumann(2011)]%
        {neumann2011efficiently}
\bibfield{author}{\bibinfo{person}{Thomas Neumann}.} \bibinfo{year}{2011}\natexlab{}.
\newblock \showarticletitle{Efficiently compiling efficient query plans for modern hardware}.
\newblock \bibinfo{journal}{\emph{Proceedings of the VLDB Endowment}} \bibinfo{volume}{4}, \bibinfo{number}{9} (\bibinfo{year}{2011}), \bibinfo{pages}{539--550}.
\newblock


\bibitem[Papadopoulos et~al\mbox{.}(2016)]%
        {PapadopoulosDMM16}
\bibfield{author}{\bibinfo{person}{Stavros Papadopoulos}, \bibinfo{person}{Kushal Datta}, \bibinfo{person}{Samuel Madden}, {and} \bibinfo{person}{Timothy~G. Mattson}.} \bibinfo{year}{2016}\natexlab{}.
\newblock \showarticletitle{The TileDB Array Data Storage Manager}.
\newblock \bibinfo{journal}{\emph{Proc. {VLDB} Endow.}} \bibinfo{volume}{10}, \bibinfo{number}{4} (\bibinfo{year}{2016}), \bibinfo{pages}{349--360}.
\newblock
\urldef\tempurl%
\url{https://doi.org/10.14778/3025111.3025117}
\showDOI{\tempurl}


\bibitem[Paszke et~al\mbox{.}(2019)]%
        {PaszkeGMLBCKLGA19}
\bibfield{author}{\bibinfo{person}{Adam Paszke} {et~al\mbox{.}}} \bibinfo{year}{2019}\natexlab{}.
\newblock \showarticletitle{{PyTorch: An Imperative Style, High- Performance Deep Learning Library}}. In \bibinfo{booktitle}{\emph{NeurIPS}}.
\newblock


\bibitem[PyTorch(2023)]%
        {backend}
\bibfield{author}{\bibinfo{person}{PyTorch}.} \bibinfo{year}{2023}\natexlab{}.
\newblock \bibinfo{title}{{PyTocrch 2.0}}.
\newblock
\newblock
\urldef\tempurl%
\url{https://pytorch.org/get-started/pytorch-2.0}
\showURL{%
\tempurl}


\bibitem[Raina et~al\mbox{.}(2009)]%
        {raina2009large}
\bibfield{author}{\bibinfo{person}{Rajat Raina}, \bibinfo{person}{Anand Madhavan}, {and} \bibinfo{person}{Andrew~Y Ng}.} \bibinfo{year}{2009}\natexlab{}.
\newblock \showarticletitle{Large-scale deep unsupervised learning using graphics processors}. In \bibinfo{booktitle}{\emph{Proceedings of the 26th annual international conference on machine learning}}. ACM, \bibinfo{pages}{873--880}.
\newblock


\bibitem[Rajbhandari et~al\mbox{.}(2021)]%
        {rajbhandari2021zero}
\bibfield{author}{\bibinfo{person}{Samyam Rajbhandari}, \bibinfo{person}{Olatunji Ruwase}, \bibinfo{person}{Jeff Rasley}, \bibinfo{person}{Shaden Smith}, {and} \bibinfo{person}{Yuxiong He}.} \bibinfo{year}{2021}\natexlab{}.
\newblock \showarticletitle{Zero-infinity: Breaking the gpu memory wall for extreme scale deep learning}. In \bibinfo{booktitle}{\emph{Proceedings of the International Conference for High Performance Computing, Networking, Storage and Analysis}}. \bibinfo{pages}{1--14}.
\newblock


\bibitem[Rocklin(2015)]%
        {rocklin2015dask}
\bibfield{author}{\bibinfo{person}{Matthew Rocklin}.} \bibinfo{year}{2015}\natexlab{}.
\newblock \showarticletitle{Dask: Parallel computation with blocked algorithms and task scheduling}. In \bibinfo{booktitle}{\emph{Proceedings of the 14th python in science conference}}, Vol.~\bibinfo{volume}{126}. Citeseer.
\newblock


\bibitem[Schaarschmidt et~al\mbox{.}(2021)]%
        {schaarschmidt2021automap}
\bibfield{author}{\bibinfo{person}{Michael Schaarschmidt}, \bibinfo{person}{Dominik Grewe}, \bibinfo{person}{Dimitrios Vytiniotis}, \bibinfo{person}{Adam Paszke}, \bibinfo{person}{Georg~Stefan Schmid}, \bibinfo{person}{Tamara Norman}, \bibinfo{person}{James Molloy}, \bibinfo{person}{Jonathan Godwin}, \bibinfo{person}{Norman~Alexander Rink}, {and} \bibinfo{person}{Vinod Nair}.} \bibinfo{year}{2021}\natexlab{}.
\newblock \showarticletitle{Automap: Towards Ergonomic Automated Parallelism for ML Models}.
\newblock \bibinfo{journal}{\emph{arXiv preprint arXiv:2112.02958}} (\bibinfo{year}{2021}).
\newblock


\bibitem[Schleich et~al\mbox{.}(2023)]%
        {schleich2023optimizing}
\bibfield{author}{\bibinfo{person}{Maximilian Schleich}, \bibinfo{person}{Amir Shaikhha}, {and} \bibinfo{person}{Dan Suciu}.} \bibinfo{year}{2023}\natexlab{}.
\newblock \showarticletitle{Optimizing Tensor Programs on Flexible Storage}.
\newblock \bibinfo{journal}{\emph{Proceedings of the ACM on Management of Data}} \bibinfo{volume}{1}, \bibinfo{number}{1} (\bibinfo{year}{2023}), \bibinfo{pages}{1--27}.
\newblock


\bibitem[Sch{\"{u}}le et~al\mbox{.}(2022)]%
        {SchuleGK022}
\bibfield{author}{\bibinfo{person}{Maximilian~E. Sch{\"{u}}le}, \bibinfo{person}{Tobias G{\"{o}}tz}, \bibinfo{person}{Alfons Kemper}, {and} \bibinfo{person}{Thomas Neumann}.} \bibinfo{year}{2022}\natexlab{}.
\newblock \showarticletitle{ArrayQL Integration into Code-Generating Database Systems}. In \bibinfo{booktitle}{\emph{{EDBT}}}.
\newblock
\urldef\tempurl%
\url{https://doi.org/10.5441/002/edbt.2022.04}
\showDOI{\tempurl}


\bibitem[Shamis et~al\mbox{.}(2015)]%
        {shamis2015ucx}
\bibfield{author}{\bibinfo{person}{Pavel Shamis}, \bibinfo{person}{Manjunath~Gorentla Venkata}, \bibinfo{person}{M~Graham Lopez}, \bibinfo{person}{Matthew~B Baker}, \bibinfo{person}{Oscar Hernandez}, \bibinfo{person}{Yossi Itigin}, \bibinfo{person}{Mike Dubman}, \bibinfo{person}{Gilad Shainer}, \bibinfo{person}{Richard~L Graham}, \bibinfo{person}{Liran Liss}, {et~al\mbox{.}}} \bibinfo{year}{2015}\natexlab{}.
\newblock \showarticletitle{UCX: an open source framework for HPC network APIs and beyond}. In \bibinfo{booktitle}{\emph{2015 IEEE 23rd Annual Symposium on High-Performance Interconnects}}. IEEE, \bibinfo{pages}{40--43}.
\newblock


\bibitem[Shazeer et~al\mbox{.}(2018)]%
        {shazeer2018mesh}
\bibfield{author}{\bibinfo{person}{Noam Shazeer}, \bibinfo{person}{Youlong Cheng}, \bibinfo{person}{Niki Parmar}, \bibinfo{person}{Dustin Tran}, \bibinfo{person}{Ashish Vaswani}, \bibinfo{person}{Penporn Koanantakool}, \bibinfo{person}{Peter Hawkins}, \bibinfo{person}{HyoukJoong Lee}, \bibinfo{person}{Mingsheng Hong}, {and} \bibinfo{person}{Cliff Young}.} \bibinfo{year}{2018}\natexlab{}.
\newblock \showarticletitle{Mesh-tensorflow: Deep learning for supercomputers}.
\newblock \bibinfo{journal}{\emph{arXiv preprint arXiv:1811.02084}} (\bibinfo{year}{2018}).
\newblock


\bibitem[Sheng et~al\mbox{.}(2023)]%
        {sheng2023flexgen}
\bibfield{author}{\bibinfo{person}{Ying Sheng}, \bibinfo{person}{Lianmin Zheng}, \bibinfo{person}{Binhang Yuan}, \bibinfo{person}{Zhuohan Li}, \bibinfo{person}{Max Ryabinin}, \bibinfo{person}{Beidi Chen}, \bibinfo{person}{Percy Liang}, \bibinfo{person}{Christopher R{\'e}}, \bibinfo{person}{Ion Stoica}, {and} \bibinfo{person}{Ce Zhang}.} \bibinfo{year}{2023}\natexlab{}.
\newblock \showarticletitle{Flexgen: High-throughput generative inference of large language models with a single gpu}. In \bibinfo{booktitle}{\emph{International Conference on Machine Learning}}. PMLR, \bibinfo{pages}{31094--31116}.
\newblock


\bibitem[Shoeybi et~al\mbox{.}(2019)]%
        {shoeybi2019megatron}
\bibfield{author}{\bibinfo{person}{Mohammad Shoeybi}, \bibinfo{person}{Mostofa Patwary}, \bibinfo{person}{Raul Puri}, \bibinfo{person}{Patrick LeGresley}, \bibinfo{person}{Jared Casper}, {and} \bibinfo{person}{Bryan Catanzaro}.} \bibinfo{year}{2019}\natexlab{}.
\newblock \showarticletitle{Megatron-lm: Training multi-billion parameter language models using model parallelism}.
\newblock \bibinfo{journal}{\emph{arXiv preprint arXiv:1909.08053}} (\bibinfo{year}{2019}).
\newblock


\bibitem[Solomonik and Demmel(2011)]%
        {solomonik2011communication}
\bibfield{author}{\bibinfo{person}{Edgar Solomonik} {and} \bibinfo{person}{James Demmel}.} \bibinfo{year}{2011}\natexlab{}.
\newblock \showarticletitle{Communication-optimal parallel 2.5 D matrix multiplication and LU factorization algorithms}. In \bibinfo{booktitle}{\emph{European Conference on Parallel Processing}}. Springer, \bibinfo{pages}{90--109}.
\newblock


\bibitem[Sompolski et~al\mbox{.}(2011)]%
        {sompolski2011vectorization}
\bibfield{author}{\bibinfo{person}{Juliusz Sompolski}, \bibinfo{person}{Marcin Zukowski}, {and} \bibinfo{person}{Peter Boncz}.} \bibinfo{year}{2011}\natexlab{}.
\newblock \showarticletitle{Vectorization vs. compilation in query execution}. In \bibinfo{booktitle}{\emph{Proceedings of the Seventh International Workshop on Data Management on New Hardware}}. \bibinfo{pages}{33--40}.
\newblock


\bibitem[Tang et~al\mbox{.}(2023)]%
        {pmlr-v202-tang23a}
\bibfield{author}{\bibinfo{person}{Yuxin Tang}, \bibinfo{person}{Zhimin Ding}, \bibinfo{person}{Dimitrije Jankov}, \bibinfo{person}{Binhang Yuan}, \bibinfo{person}{Daniel Bourgeois}, {and} \bibinfo{person}{Chris Jermaine}.} \bibinfo{year}{2023}\natexlab{}.
\newblock \showarticletitle{Auto-Differentiation of Relational Computations for Very Large Scale Machine Learning}. In \bibinfo{booktitle}{\emph{Proceedings of the 40th International Conference on Machine Learning}} \emph{(\bibinfo{series}{Proceedings of Machine Learning Research}, Vol.~\bibinfo{volume}{202})}, \bibfield{editor}{\bibinfo{person}{Andreas Krause}, \bibinfo{person}{Emma Brunskill}, \bibinfo{person}{Kyunghyun Cho}, \bibinfo{person}{Barbara Engelhardt}, \bibinfo{person}{Sivan Sabato}, {and} \bibinfo{person}{Jonathan Scarlett}} (Eds.). \bibinfo{publisher}{PMLR}, \bibinfo{pages}{33581--33598}.
\newblock
\urldef\tempurl%
\url{https://proceedings.mlr.press/v202/tang23a.html}
\showURL{%
\tempurl}


\bibitem[Touvron et~al\mbox{.}(2023)]%
        {touvron2023llama}
\bibfield{author}{\bibinfo{person}{Hugo Touvron}, \bibinfo{person}{Thibaut Lavril}, \bibinfo{person}{Gautier Izacard}, \bibinfo{person}{Xavier Martinet}, \bibinfo{person}{Marie-Anne Lachaux}, \bibinfo{person}{Timoth{\'e}e Lacroix}, \bibinfo{person}{Baptiste Rozi{\`e}re}, \bibinfo{person}{Naman Goyal}, \bibinfo{person}{Eric Hambro}, \bibinfo{person}{Faisal Azhar}, {et~al\mbox{.}}} \bibinfo{year}{2023}\natexlab{}.
\newblock \showarticletitle{Llama: Open and efficient foundation language models}.
\newblock \bibinfo{journal}{\emph{arXiv preprint arXiv:2302.13971}} (\bibinfo{year}{2023}).
\newblock


\bibitem[Vasilache et~al\mbox{.}(2018)]%
        {vasilache2018tensor}
\bibfield{author}{\bibinfo{person}{Nicolas Vasilache}, \bibinfo{person}{Oleksandr Zinenko}, \bibinfo{person}{Theodoros Theodoridis}, \bibinfo{person}{Priya Goyal}, \bibinfo{person}{Zachary DeVito}, \bibinfo{person}{William~S Moses}, \bibinfo{person}{Sven Verdoolaege}, \bibinfo{person}{Andrew Adams}, {and} \bibinfo{person}{Albert Cohen}.} \bibinfo{year}{2018}\natexlab{}.
\newblock \showarticletitle{Tensor comprehensions: Framework-agnostic high-performance machine learning abstractions}.
\newblock \bibinfo{journal}{\emph{arXiv preprint arXiv:1802.04730}} (\bibinfo{year}{2018}).
\newblock


\bibitem[Vaswani et~al\mbox{.}(2017)]%
        {vaswani2017attention}
\bibfield{author}{\bibinfo{person}{Ashish Vaswani}, \bibinfo{person}{Noam Shazeer}, \bibinfo{person}{Niki Parmar}, \bibinfo{person}{Jakob Uszkoreit}, \bibinfo{person}{Llion Jones}, \bibinfo{person}{Aidan~N Gomez}, \bibinfo{person}{{\L}ukasz Kaiser}, {and} \bibinfo{person}{Illia Polosukhin}.} \bibinfo{year}{2017}\natexlab{}.
\newblock \showarticletitle{Attention is all you need}. In \bibinfo{booktitle}{\emph{Advances in Neural Information Processing Systems}}. \bibinfo{pages}{6000--6010}.
\newblock


\bibitem[Xu et~al\mbox{.}(2021)]%
        {xu2021gspmd}
\bibfield{author}{\bibinfo{person}{Yuanzhong Xu}, \bibinfo{person}{HyoukJoong Lee}, \bibinfo{person}{Dehao Chen}, \bibinfo{person}{Blake Hechtman}, \bibinfo{person}{Yanping Huang}, \bibinfo{person}{Rahul Joshi}, \bibinfo{person}{Maxim Krikun}, \bibinfo{person}{Dmitry Lepikhin}, \bibinfo{person}{Andy Ly}, {and} \bibinfo{person}{Marcello Maggioni}.} \bibinfo{year}{2021}\natexlab{}.
\newblock \showarticletitle{{GSPMD}: General and Scalable Parallelization for ML Computation Graphs}.
\newblock \bibinfo{journal}{\emph{arXiv preprint arXiv:2105.04663}} (\bibinfo{year}{2021}).
\newblock


\bibitem[Yuan et~al\mbox{.}(2021)]%
        {yuan2021tensor}
\bibfield{author}{\bibinfo{person}{Binhang Yuan}, \bibinfo{person}{Dimitrije Jankov}, \bibinfo{person}{Jia Zou}, \bibinfo{person}{Yuxin Tang}, \bibinfo{person}{Daniel Bourgeois}, {and} \bibinfo{person}{Chris Jermaine}.} \bibinfo{year}{2021}\natexlab{}.
\newblock \showarticletitle{Tensor Relational Algebra for Distributed Machine Learning System Design}.
\newblock  (\bibinfo{year}{2021}).
\newblock


\bibitem[Zhang et~al\mbox{.}(1990)]%
        {zhang1990efficient}
\bibfield{author}{\bibinfo{person}{Xiru Zhang}, \bibinfo{person}{Michael Mckenna}, \bibinfo{person}{Jill~P Mesirov}, {and} \bibinfo{person}{David~L Waltz}.} \bibinfo{year}{1990}\natexlab{}.
\newblock \showarticletitle{An efficient implementation of the back-propagation algorithm on the connection machine {CM-2}}. In \bibinfo{booktitle}{\emph{Advances in neural information processing systems}}. \bibinfo{pages}{801--809}.
\newblock


\bibitem[Zheng et~al\mbox{.}(2022)]%
        {zheng2022alpa}
\bibfield{author}{\bibinfo{person}{Lianmin Zheng}, \bibinfo{person}{Zhuohan Li}, \bibinfo{person}{Hao Zhang}, \bibinfo{person}{Yonghao Zhuang}, \bibinfo{person}{Zhifeng Chen}, \bibinfo{person}{Yanping Huang}, \bibinfo{person}{Yida Wang}, \bibinfo{person}{Yuanzhong Xu}, \bibinfo{person}{Danyang Zhuo}, \bibinfo{person}{Eric~P Xing}, {et~al\mbox{.}}} \bibinfo{year}{2022}\natexlab{}.
\newblock \showarticletitle{Alpa: Automating inter-and $\{$Intra-Operator$\}$ parallelism for distributed deep learning}. In \bibinfo{booktitle}{\emph{16th USENIX Symposium on Operating Systems Design and Implementation (OSDI 22)}}. \bibinfo{pages}{559--578}.
\newblock


\end{thebibliography}

\end{document}